\documentclass{emulateapj}
\usepackage{graphicx}
\usepackage{natbib}
\usepackage{appendix}
\slugcomment{To appear in ApJ., XX.}
\shorttitle{Rate of SN\,II in Galaxy Clusters}
\shortauthors{Graham et al.}
\begin{document}

\title{The Type II Supernova Rate in $\lowercase{z}\sim0.1$ Galaxy Clusters \\ from the Multi-Epoch Nearby Cluster Survey}

\author{
M.~L. Graham\altaffilmark{1,2},
D.~J. Sand\altaffilmark{1,2,3},
C.~J. Bildfell\altaffilmark{4},
C.~J. Pritchet\altaffilmark{4},
D. Zaritsky\altaffilmark{5},
H. Hoekstra\altaffilmark{6},
D.~W. Just\altaffilmark{5},
S. Herbert-Fort\altaffilmark{5},
S. Sivanandam\altaffilmark{7},
R.~J. Foley $\!$\altaffilmark{8,9}}

\altaffiltext{1}{Las Cumbres Observatory Global Telescope Network, 6740
Cortona Drive, Suite 102, Santa Barbara, CA 93117, USA}
\altaffiltext{2}{Department of Physics, Broida Hall, University of
California, Santa Barbara, CA 93106, USA}
\altaffiltext{3}{Harvard Center for Astrophysics and Las Cumbres Observatory Global Telescope Network Fellow}
\altaffiltext{4}{Department of Physics and Astronomy, University of
Victoria, PO Box 3055, STN CSC, Victoria BC V8W 3P6, Canada}
\altaffiltext{5}{Steward Observatory, University of Arizona, Tucson AZ 85721}
\altaffiltext{6}{Leiden Observatory, Leiden University, Niels Bohrweg 2, NL-2333 CA Leiden, The Netherlands}
\altaffiltext{7}{Dunlap Fellow, Dunlap Institute for Astronomy and Astrophysics, 50 St. George St., Toronto, ON Canada M5S 3H4}
\altaffiltext{8}{Harvard-Smithsonian Center for Astrophysics, 60 Garden Street, Cambridge, MA 02138, USA}
\altaffiltext{9}{Clay Fellow.}

\begin{abstract}
We present 7 spectroscopically confirmed Type II cluster supernovae (SNe\,II) discovered in the Multi-Epoch Nearby Cluster Survey, a supernova survey targeting 57 low redshift $0.05<z<0.15$ galaxy clusters with the Canada-France-Hawaii Telescope. We find the rate of Type II supernovae within $R_{200}$ of $z\sim0.1$ galaxy clusters to be $0.026_{-0.018}^{+0.085}$(stat)$_{-0.001}^{+0.003}$(sys) SNuM. Surprisingly, one SN\,II is in a red sequence host galaxy that shows no clear evidence of recent star formation. This is unambiguous evidence in support of ongoing, low-level star formation in at least some cluster elliptical galaxies, and illustrates that galaxies that appear to be quiescent cannot be assumed to host only Type Ia SNe. Based on this single SN\,II we make the first measurement of the SN\,II rate in red sequence galaxies, and find it to be $0.007_{-0.007}^{+0.014}$(stat)$_{-0.001}^{+0.009}$(sys) SNuM. We also make the first derivation of cluster specific star formation rates (sSFR) from cluster SN\,II rates. We find that for all galaxy types, the sSFR is $5.1_{-3.1}^{+15.8}$(stat)$\pm$0.9(sys) $\rm M_{\odot} \ yr^{-1} \ (10^{12}M_{\odot})^{-1}$, and for red sequence galaxies only, it is $2.0_{-0.9}^{+4.2}$(stat)$\pm$0.4(sys) $\rm M_{\odot} \ yr^{-1} \ (10^{12}M_{\odot})^{-1}$. These values agree with SFRs measured from infrared and ultraviolet photometry, and H$\alpha$ emission from optical spectroscopy. Additionally, we use the SFR derived from our SNII rate to show that although a small fraction of cluster Type Ia SNe may originate in the young stellar population and experience a short delay time, these results do not preclude the use of cluster SN\,Ia rates to derive the late-time delay time distribution for SNe\,Ia.
\end{abstract}

\keywords{supernovae: general --- galaxies: clusters}

\section{Introduction}\label{s:intro}

The rates and properties of Type II supernovae (SNe\,II) indicate they are explosions induced by the collapse of iron cores in stars of initial masses $8<M\lesssim20$ $\rm M_{\odot}$ \citep{Smartt09}. Such massive progenitors have been directly confirmed in pre-explosion images for several SNe\,II \citep{Li07a, galyam07, Smartt09, ER11, Maund11}. Stars of initial mass $>8$ $\rm M_{\odot}$ explode as SNe with a delay time, the time between star formation and explosion, of $\lesssim30$ Myr (e.g. Henyey et al. 1959). Due to this relatively short delay time the SN\,II rate, $SNR_{II}$, is a direct indication of the current star formation rate, $SFR$ (Botticella et al. 2012). Since SNe\,II are bright, using $SNR_{II}$ to identify very low levels of star formation (SF) can be advantageous to using ultraviolet (UV) photometry or optical spectroscopy. For example, small amounts of UV light may be undetectable in a luminous elliptical, and a little H$\alpha$ emission in an elliptical can be overwhelmed by an active galactic nucleus. Also, at low redshift UV photometry can be an ambiguous SF tracer because of contributions from blue horizontal branch stars (the UV upturn). The $SNR_{II}$ is thus an especially valuable SFR proxy in rich galaxy clusters, which are mainly composed of luminous elliptical galaxies. 

Evidence for low levels of SF has recently been detected in low redshift cluster galaxies from optical spectra \citep{wings} and infrared photometry \citep{Chung11}, and in low redshift field red sequence galaxies from UV photometry \citep{kaviraj2010}. Despite this, SNe\,II have rarely been observed in early-type galaxies where the bulk of the stellar mass is in old stellar populations \citep{Hako08}. As a result, the $SNR_{II}$ in field ellipticals and cluster galaxies is not well constrained. Table \ref{table:IIrates} presents the current literature values for $SNR_{II}$. The Lick Observatory Supernova Search (LOSS; Leaman et al. 2011) placed an upper limit on $SNR_{II}$ in field ellipticals, and provided the first measurement of $SNR_{II}$ in field S0 galaxies \citep{Li11b}. Five SN surveys were combined by Mannucci et al. (2008; hereafter M08), who measured the $SNR_{II}$ in galaxy clusters to be half the rate in field galaxies, but well above the upper limits of field ellipticals. However, the surveys complied by M08 were biased towards the most massive cluster members, and their SNe were not all spectroscopically classified. In this paper we present the $SNR_{II}$ measurement from our large, complete, well characterized cluster SN survey at low redshifts, and make the first comparison of SN\,II-derived cluster SFR to the recently detected low levels of SF from optical, IR, and UV data.

In contrast to SNe\,II, SNe\,Ia occur in both young and old stellar populations \citep{mann05,sb05}. Ongoing star formation in galaxy clusters is a concern when using the cluster $SNR_{Ia}$ to constrain the slope of the SN\,Ia delay time distribution (DTD). The two leading scenarios for the SN\,Ia progenitor are a carbon-oxygen white dwarf accreting material from a main sequence or red giant star (single degenerate), or accreting from or merging with another white dwarf (double degenerate). Each occur over different timescales and predict distinctive DTDs at late times; i.e. the double degenerate scenario predicts more delayed SNe\,Ia. The colors of cluster red sequence galaxies indicate that their star formation was truncated at high redshift, and they have evolved passively since (e.g. Stanford et al. 1998; Eisenhardt et al. 2008). Based on this, it is assumed that cluster SNe\,Ia have all experienced long delay times, and that the cluster $SNR_{Ia}(z)$ can constrain the late-time DTD (i.e. Maoz et al. 2010). However, the presence of ongoing star formation suggests the cluster population may be contaminated by short-delay SNe\,Ia. In this paper we use our SN\,II-derived cluster $SFR$ to evaluate this possibility.

The Multi-Epoch Nearby Cluster Survey (MENeaCS) surveyed 57 low redshift $0.05<z<0.15$ galaxy clusters for two years with the Canada-France-Hawaii Telescope (CFHT). Within the virial radius, $R_{200}$, of our cluster sample, we spectroscopically classified  7 cluster SNe\,II and 23 cluster Type Ia supernovae. This paper is part of a series based on MENeaCS. In Sand et al. (2011; hereafter S11) we use the relative number of hosted and hostless SNe\,Ia to determine the mass fraction of intracluster stars. In Sand et al. (2012; hereafter S12), we measure the cluster $SNR_{Ia}$ from MENeaCS, and combine it with published SN\,Ia rates between $0.02<z<1.12$ to constrain the slope of the SN\,Ia delay-time distribution. Two additional MENeaCS papers are nearing publication: one showing evolution in the cluster dwarf-to-giant galaxy ratio over redshift (Bildfell et al. 2012), and one investigating the demographics of tidally disturbed galaxies in clusters (Adams et al. in preparation).

In this work we present the 7 cluster SNe\,II, including the unprecedented occurrence of a SN\,II in a red sequence galaxy. In \S~\ref{s:SNII} we describe MENeaCS, and present the photometric and spectroscopic properties of our 7 cluster SNe\,II and their hosts galaxies, including a comparison of SN\,II and SN\,Ia hosts. In \S~\ref{s:RShost} we constrain the level of ongoing star formation in our one red sequence SN\,II host galaxy using published multi-wavelength data. In \S~\ref{s:rate} we describe our SN\,II rate calculation and its uncertainties, and compare our results to published rates. In \S~\ref{s:analysis} we derive the star formation rate in clusters, and compare it to measurements from IR, UV, and spectroscopic observations of cluster galaxies. We also discuss the implications for the SN\,Ia DTD. We provide a summary of this paper in \S~\ref{s:conc}, and in all cases we use a standard flat cosmology of $(\Omega_M,\Omega_{\Lambda}) = (0.3,0.7)$ and $h=0.7$.

\section{The MENeaCS SN\,II Sample}\label{s:SNII}

MENeaCS monitored 57 rich galaxy clusters with monthly cadence for two years in the $g^\prime$- and $r^\prime$-band filters with MegaCam \citep{cfhtmegacam} at the CFHT. The survey design, cluster sample, observing strategy, real-time reductions, transient detection pipeline, photometric calibration, host identification, and spectroscopic SN classification techniques we used to discover supernovae are presented in S12. The data were reduced and searched for transients in real time, with regular spectroscopic runs scheduled each month of the survey for supernova classification with the Blue Channel Spectrograph (BCS; Schmidt et al. 1989) or Hectospec (Fabricant et al. 2005) at the MMT Observatory. Photometric calibrations to SDSS filters $g$ and $r$ in the AB magnitude system were performed using standard stars in the fields.

We spectroscopically followed up all SN candidates brighter than $m_{g}$=22.5 magnitudes, and with colors $g-r>0.8$. As described in S12, we used the publicly available Supernova Identification (SNID) routine of Blondin \& Tonry (2007) to spectroscopically classify our SNe. Cluster membership was assigned for SNe with $|v_{SN}-v_{cluster}|<3000$ $\rm km \ s^{-1} $. In total we confirmed 23 SNe\,Ia (4 of which were hostless), 7 SNe\,II associated with our clusters, and 37 background SNe. Due to the one square degree field of view of MegaCam, we are complete to $\sim R_{200}$ and can calculate SN rates within this radius. For comparison with past surveys, we also present rates within $R=1$ Mpc.

\begin{figure*}
\begin{center}
\includegraphics[width=7.5cm]{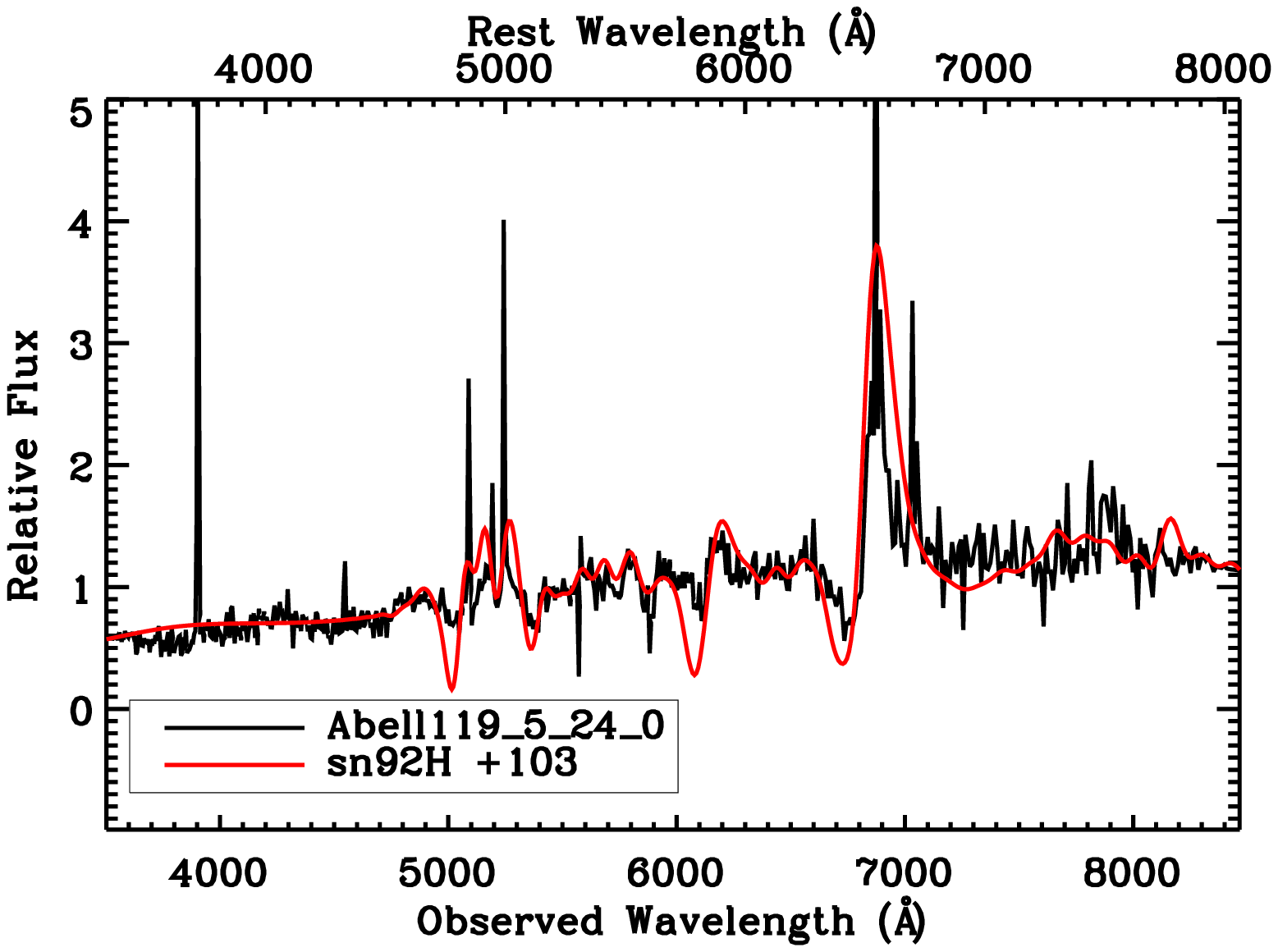}
\includegraphics[width=7.5cm]{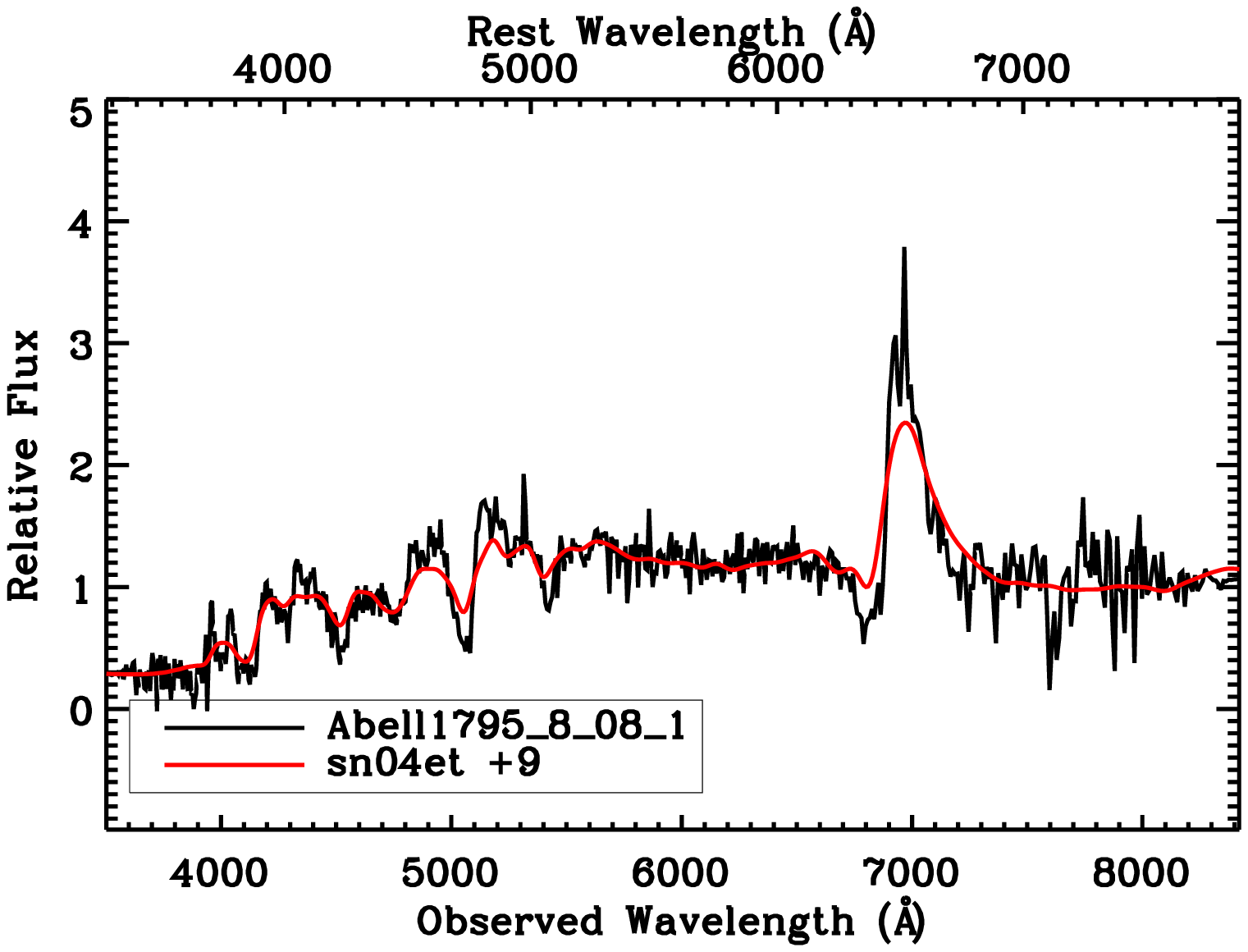}
\includegraphics[width=7.5cm]{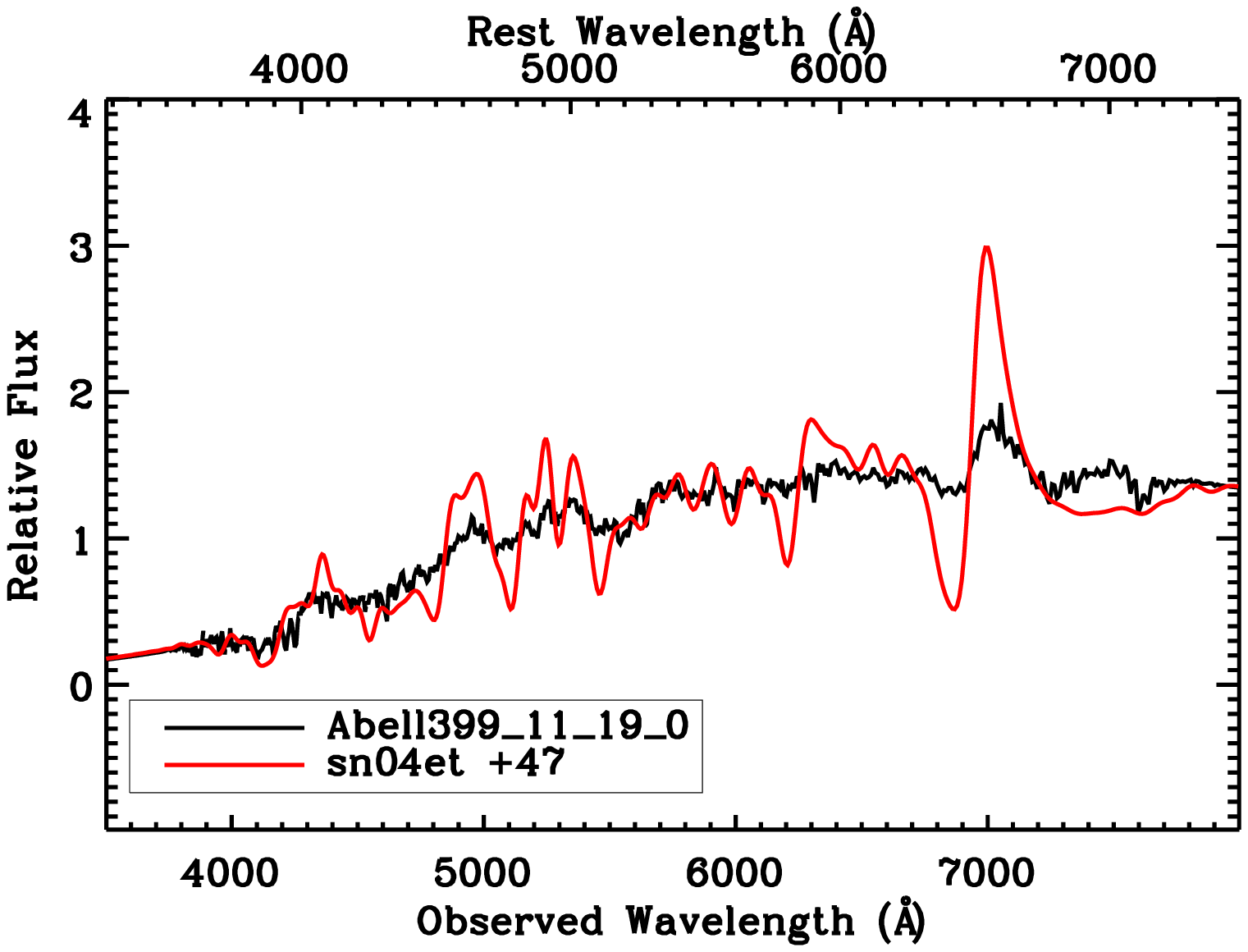}
\includegraphics[width=7.5cm]{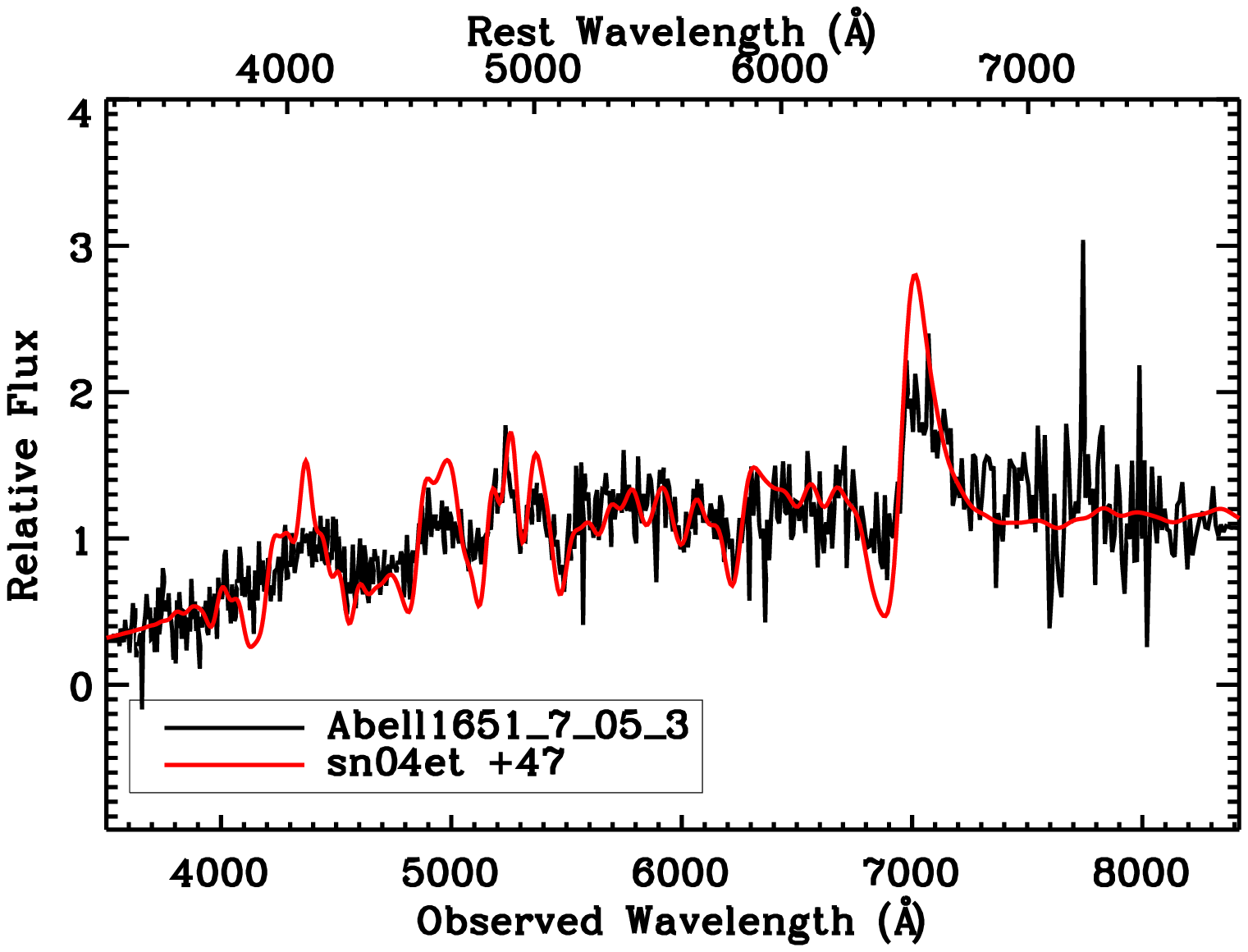}
\includegraphics[width=7.5cm]{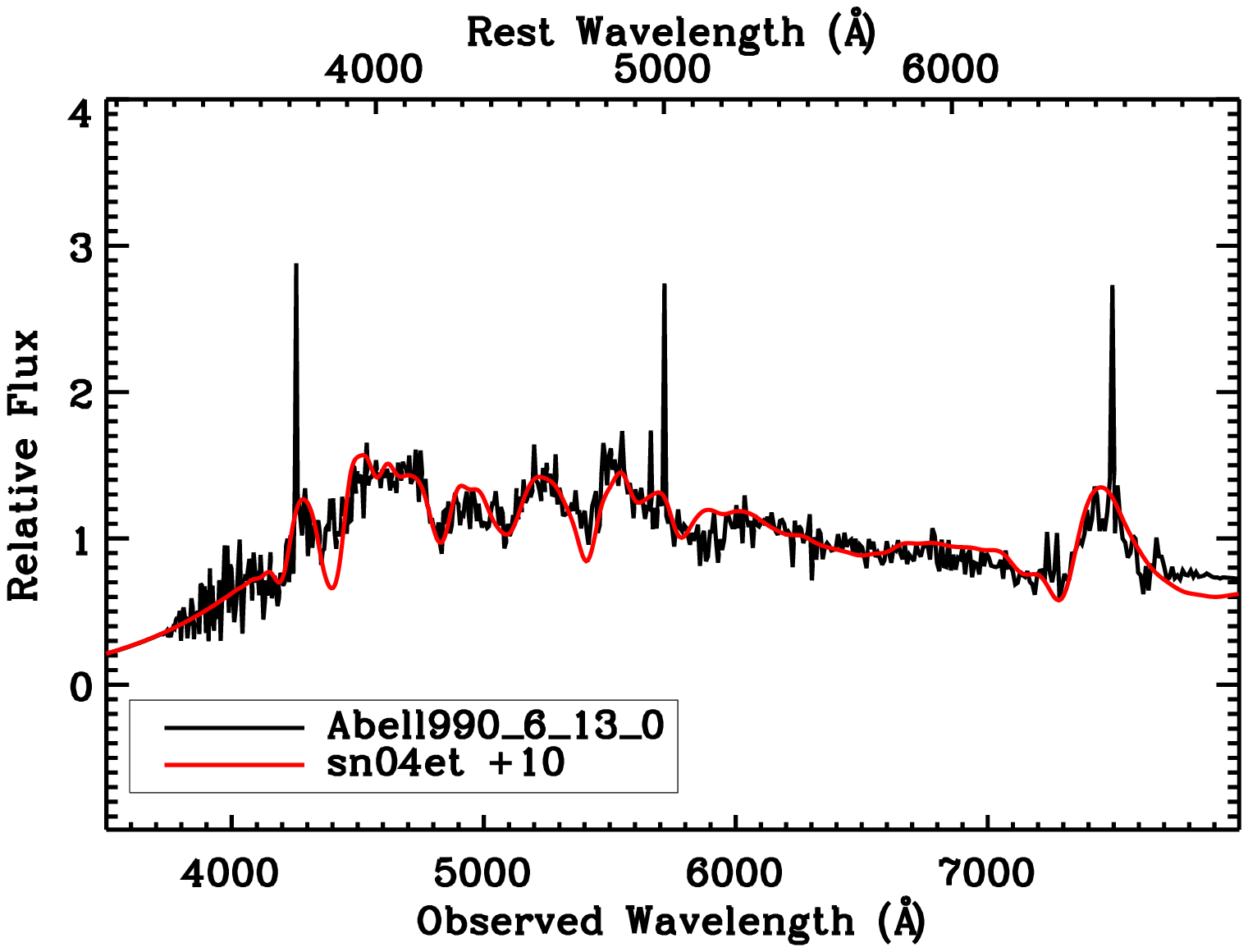}
\includegraphics[width=7.5cm]{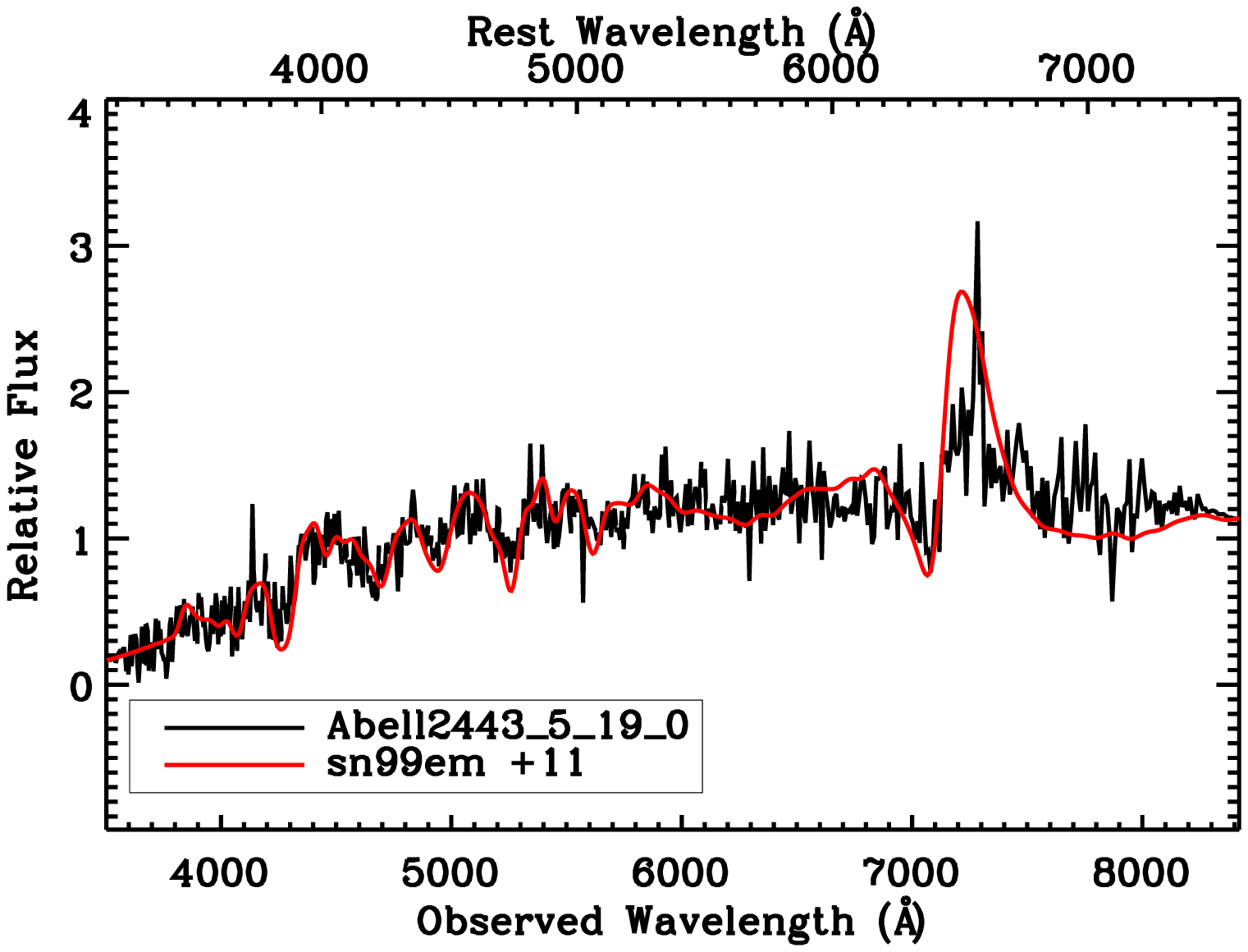}
\includegraphics[width=7.5cm]{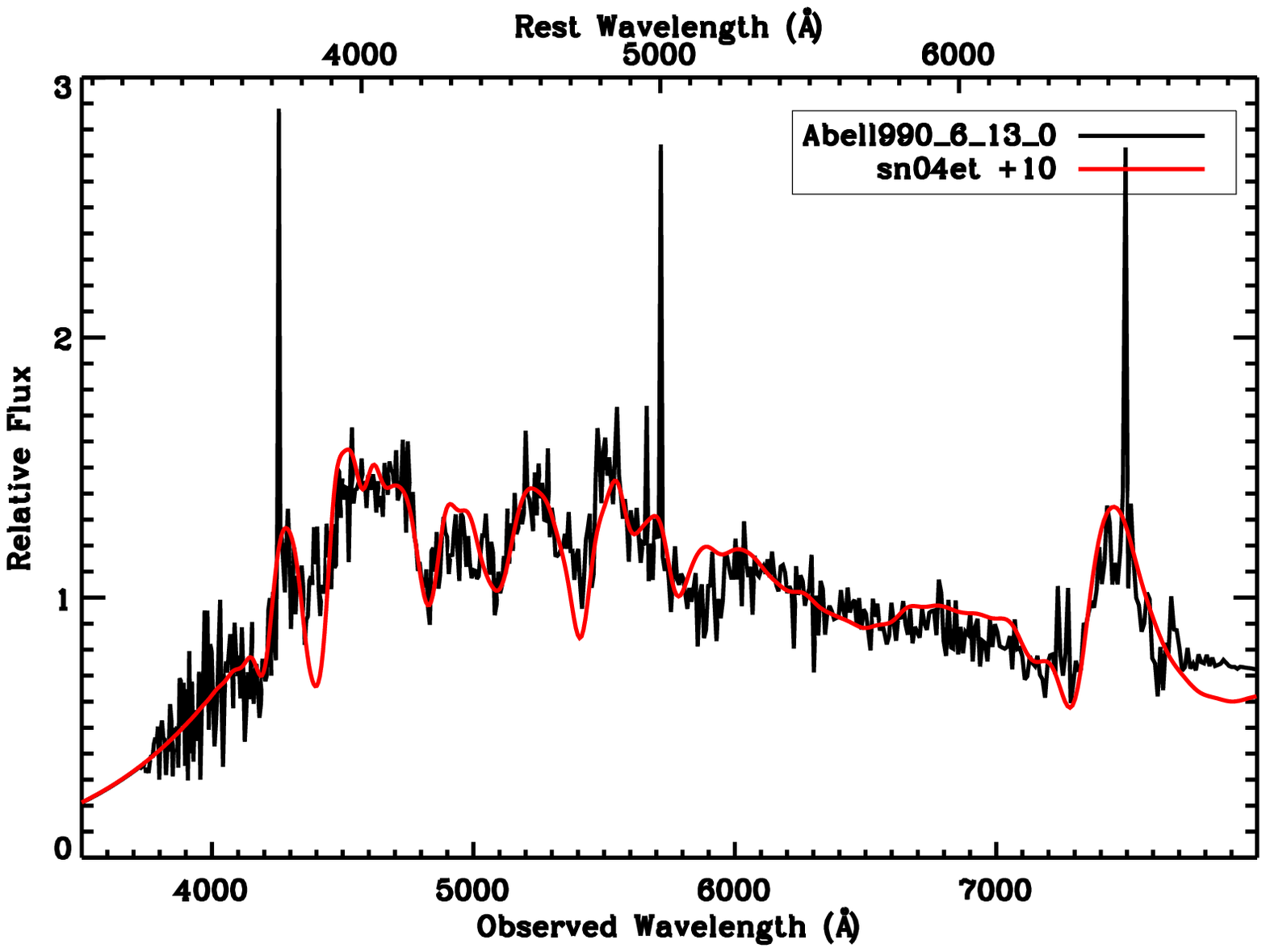}
\caption{Spectra for our cluster SNe\,II in black, with the best fitting SN\,II template spectrum from SNID in red (see Table \ref{table:CC}). \label{fig:SNset1} }
\end{center}
\end{figure*}

Table \ref{table:CC} presents the 7 SNe\,II discovered in MENeaCS clusters including their internal identification name; the UT date of spectroscopy; the telescope and instrument for follow-up; cluster redshift; galaxy redshift where available; details of the SNID best fit including redshift (and uncertainty), SN template type and name, and median phase (and the standard deviation); and finally the spectroscopic exposure time in seconds. Figure \ref{fig:SNset1} presents the classification spectra for each; all 7 are best fit with Type II plateau spectral templates. 

As discussed by Li et al. (2011a), reliable distinction between Type II subtypes requires multiple spectra {\it and} well sampled light curves. Type IIn (narrow spectral lines) are spectrally distinctive but can evolve to resemble a regular Type II. The Type IIL are spectroscopically similar to Type IIP, but have a linearly declining light curve and no plateau phase. Type IIb (broad spectral lines) can resemble normal Type II at early times, but their light curves are distinctly double-peaked. With the single epoch of spectroscopy and monthly photometric cadence of MENeaCS, we cannot confidently identify subtypes for our SNe\,II. 

\subsection{MENeaCS SN\,II Host Galaxies}\label{ss:hosts}

Photometry for MENeaCS SN\,II host galaxies is measured from SN-free deep stack images, as described in S12. Table \ref{table:hosts} presents the host galaxy details for each cluster SN\,II, including the coordinates, $r$-band magnitude, $g-r$ color, the $\Delta(g-r)_{RS}$ color offset from the cluster's red sequence, and the clustercentric radius in units of kiloparsecs and $R_{200}$. Clustercentric radius is the distance from the brightest cluster galaxy (BCG); for our 7 SNe\,II host clusters the BCGs are within 30\arcsec \ of the X-ray centers (from Chandra and ROSAT), except for Abell 2443 which has a $\sim1.5^\prime$ offset. However, even a potential $1.5^\prime$ shift could not cause any SNe\,II to cross the $\rm R_{200}$ boundary, and be included or excluded from the sample.

\begin{figure*}
\begin{center}
\includegraphics[width=8cm]{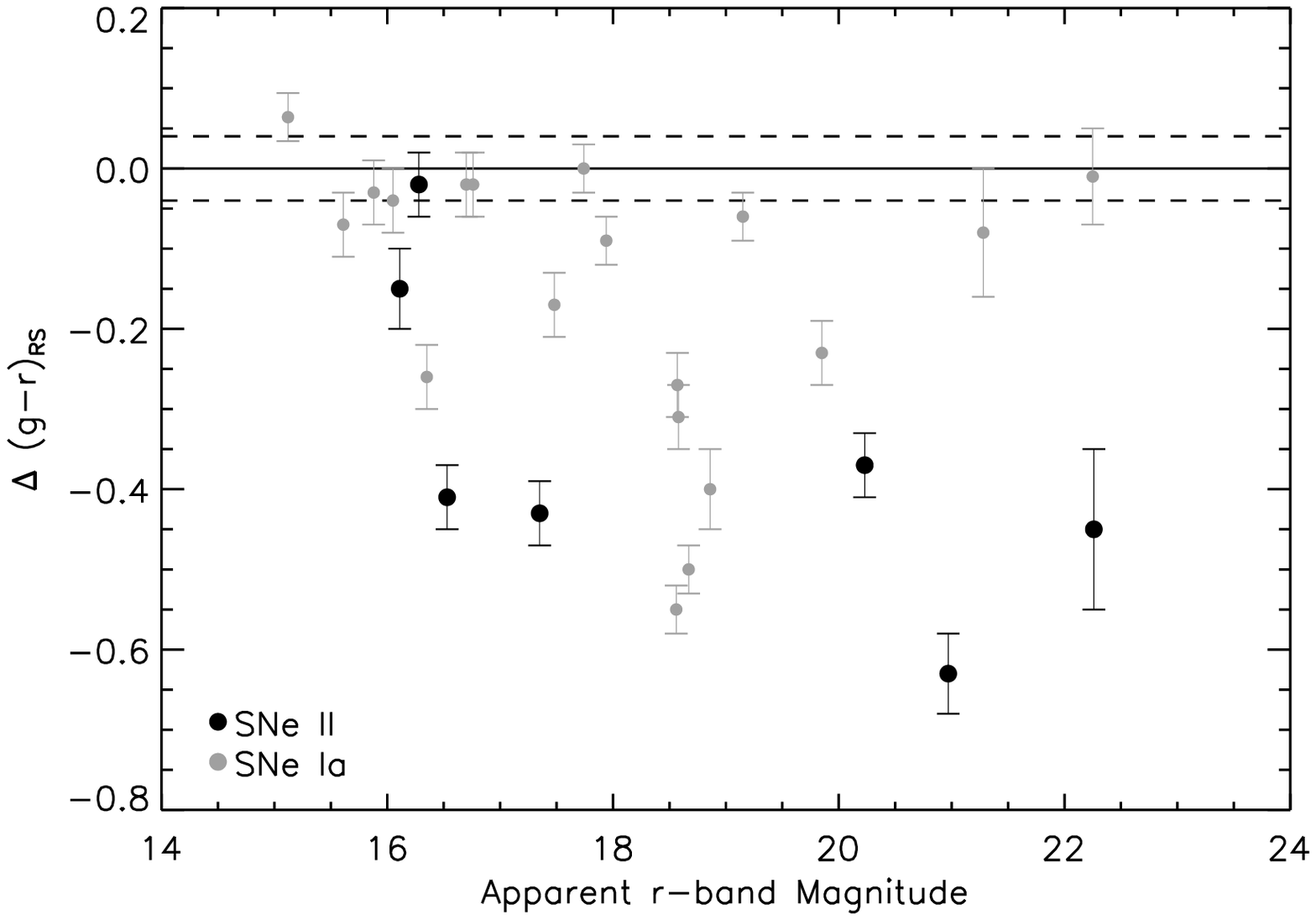}
\includegraphics[width=8cm]{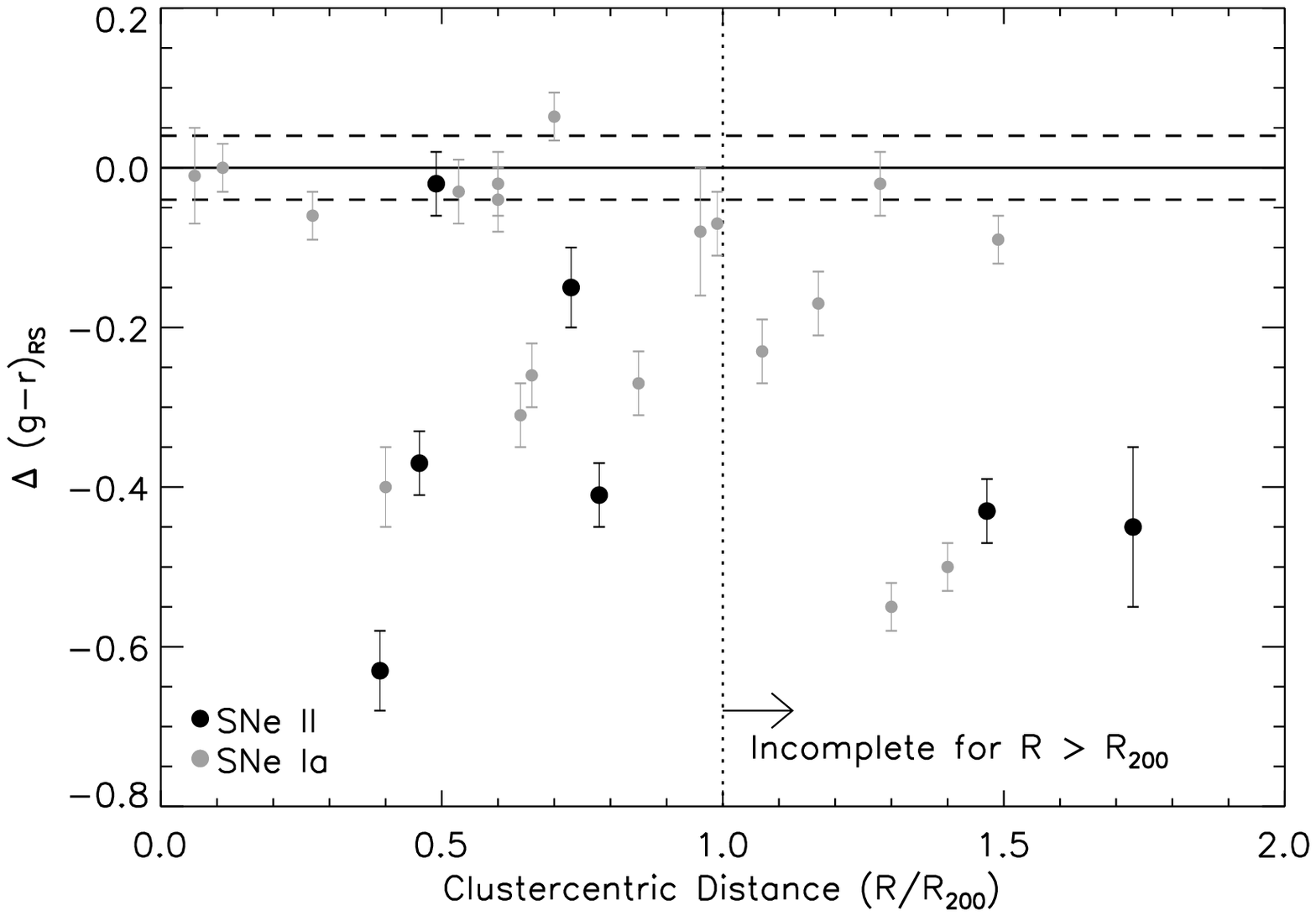}
\caption{The offset of each SN\,II (black) and SN\,Ia (gray) host galaxy from its cluster's red sequence, $\Delta (g-r)_{RS}$, versus the host galaxy $r$-band apparent magnitude (left) and the clustercentric offset in units of $R_{200}$ (right). Dashed lines represent the median scatter in the red sequence over the whole MENeaCS sample. \label{fig:RoffCC}}
\end{center}
\end{figure*}

\begin{figure*}
\begin{center}
\includegraphics[width=8cm]{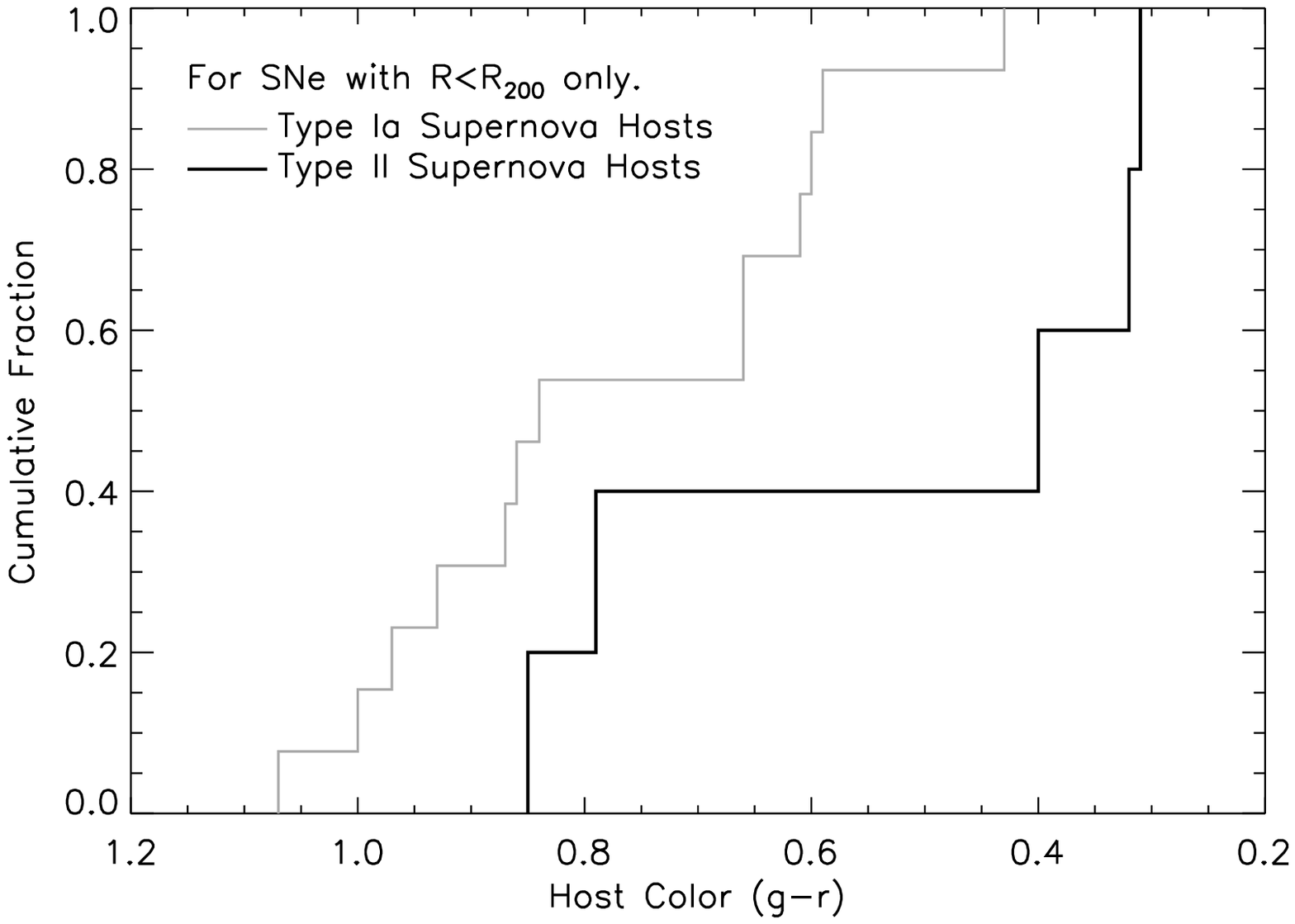}
\includegraphics[width=8cm]{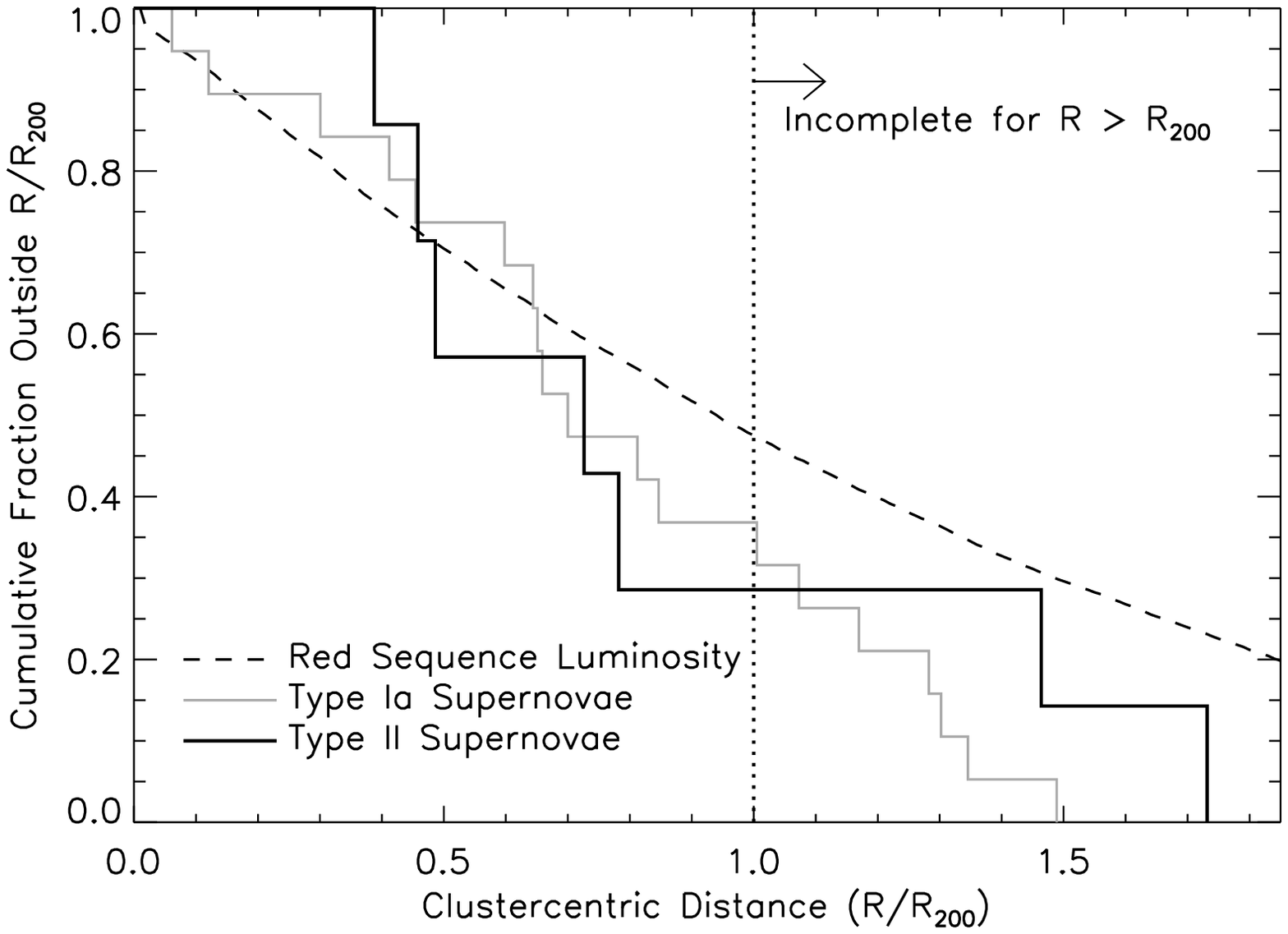}
\caption{Left: The normalized cumulative distributions of SN\,Ia (gray) and SN\,II (black) host $g-r$ colors, for SNe within $R_{200}$. A two-sided KS test finds there a $<10$\% probability that these are drawn from the same underlying distribution. Right: normalized, cumulative $g$-band luminosity in red sequence galaxies (dashed), and the cumulative number fractions of SNe\,Ia (gray) and SNe\,II (black). Within $R_{200}$, the distribution of projected clustercentric offsets for SNe\,Ia appears to ``follow the light" in the red sequence better than SNe\,II, as expected. \label{fig:cumul}}
\end{center}
\end{figure*}

Given that SN\,II progenitors are young, but SNe\,Ia occur in both young and old stellar populations, we expect the distributions of $g-r$ colors to be different for SN\,II and SN\,Ia host galaxies. In Figure \ref{fig:RoffCC} we plot the $\Delta(g-r)_{RS}$ color offset from the cluster red sequence as a function of $r$-band magnitude for SN\,Ia and SN\,II hosts. The dashed lines represent the median scatter in the red sequence for the MENeaCS sample. All host galaxies with $\Delta(g-r)_{RS}$ error bars overlapping this zone are considered to lie ``on" the red sequence, and all others lie ``off" the red sequence. Unlike the SNe\,Ia host population, SN\,II hosts lie off their clusters' red sequences -- except for the host SN\,II Abell399\_11\_19\_0, discussed in \S~\ref{s:RShost}. In Figure \ref{fig:cumul}, we show that the distributions of host $g-r$ colors are significantly different for SNe\,Ia and SNe\,II. The KS-test probability that the two samples are drawn from the same underlying distribution is low, but not negligible, at $\sim8$\%. 

Chung et al. (2011) show that the fraction of star forming galaxies increases with projected clustercentric radius within $R_{200}$. We therefore expect the radial distribution of SN\,II hosts to differ from SN\,Ia hosts, which should in turn follow the cluster luminosity profile. In Figure \ref{fig:RoffCC} we plot the $\Delta(g-r)_{RS}$ color offset from the cluster red sequence as a function of projected clustercentric radius, for both SN\,Ia and SN\,II hosts. We note that no off-RS MENeaCS hosts are observed within 0.3 $R_{200}$, consistent with the view that the star formation fraction decreases in the central regions of clusters \citep{Chung11}. Figure \ref{fig:cumul} shows the distributions of projected clustercentric radii for SN\,Ia and SN\,II hosts. The normalized, cumulative fraction of cluster g-band luminosity in red sequence galaxies is also plotted. Within $R_{200}$ it appears the number of SNe is roughly proportional to luminosity.

\section{The Red Sequence SN\,II Host} \label{s:RShost}

\begin{figure*}
\begin{center}
\includegraphics[width=7cm]{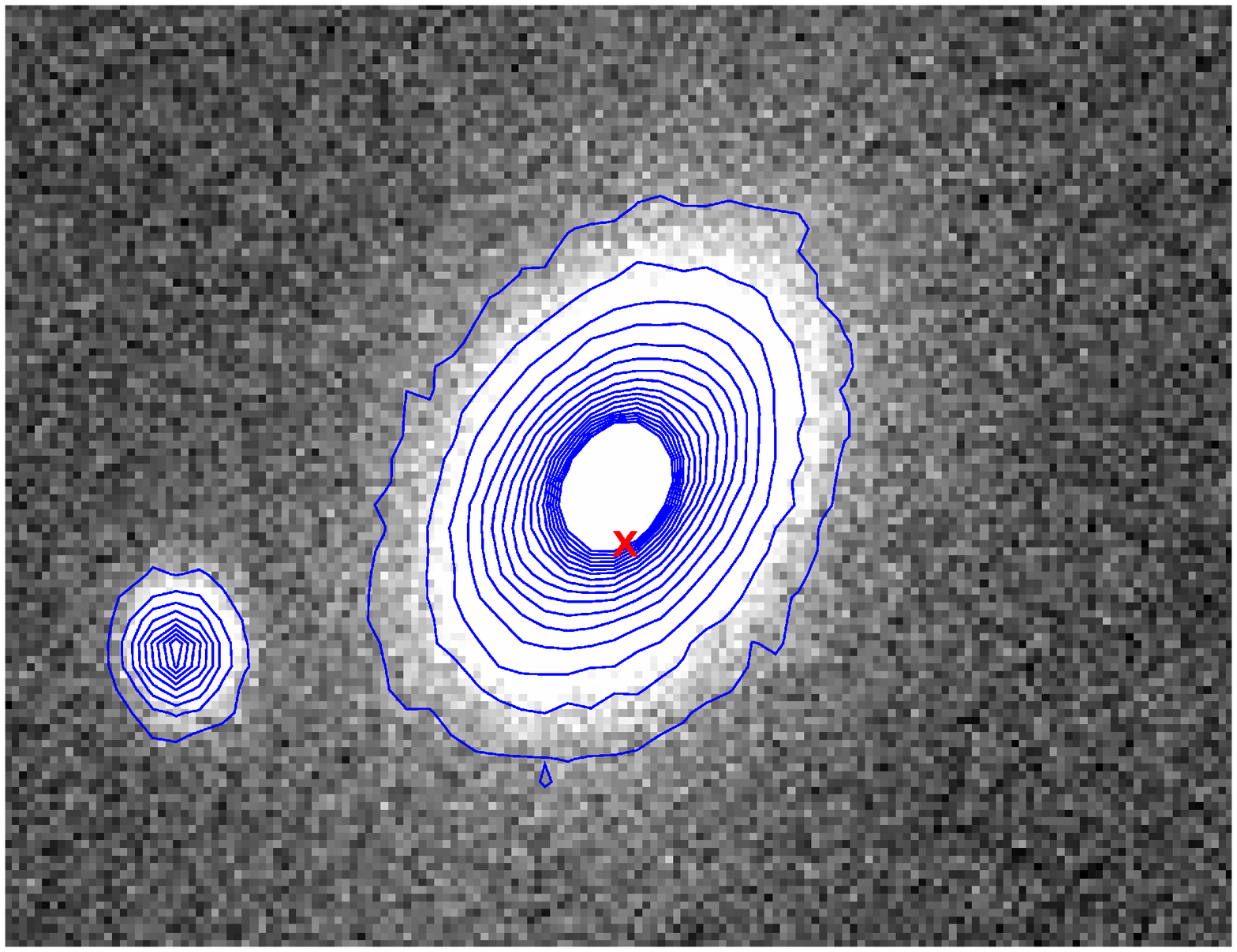}
\includegraphics[width=8cm]{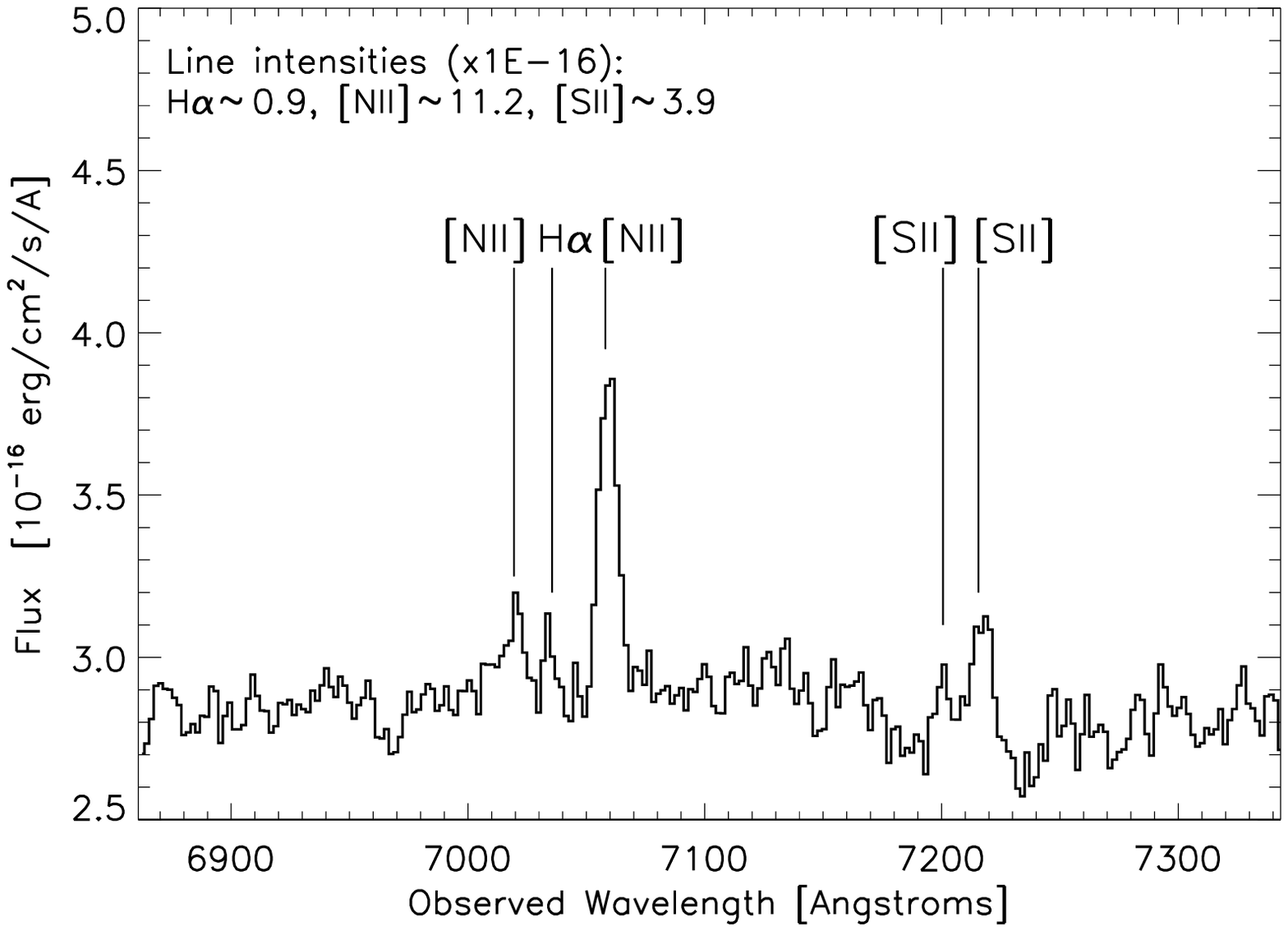}
\caption{Left: The host of SN\,II Abell399\_11\_19\_0, with the SN marked as a red cross, and smoothed isophotal contours to highlight this galaxy's elliptical morphology. Right: A partial spectrum of the red sequence host galaxy of SN\,II Abell399\_11\_19\_0, showing the relevant emission lines. Line intensities are estimated by integrating the flux over the line width. \label{fig:a399spec}}
\end{center}
\end{figure*}

Unexpectedly, one of our seven MENeaCS SNe\,II occurred in a red sequence galaxy. As we discuss below, based on its spectrum and photometry we are confident that it was a SN\,II and not a SN\,Ia. The color, magnitude, red sequence offset, and clustercentric distance for the host galaxy of Abell399\_11\_19\_0 are listed in Table \ref{table:hosts}, and are shown to be consistent with the red sequence in Figure \ref{fig:RoffCC}. In Figure \ref{fig:a399spec} we show an image of this galaxy with isophotal contours to highlight this galaxy's elliptical morphology.

When core collapse SN (CC\,SNe, Types II and Ibc) are discovered in elliptical galaxies, further inspection almost always reveals the presence of star formation. Hakobyan et al. (2008) appraised 22 elliptical galaxies hosting CC\,SNe; of them, 19 were misclassified as elliptical, and three showed evidence of mergers or interactions and were thus likely to harbor recent star formation. Suh et al. (2011) investigated the near-ultraviolet and radio properties of nine early-type CC\,SN hosts, finding clear evidence of recent star formation in all. These results are consistent with the relative youth of SN\,II progenitor stars. In this section we look for evidence of recent star formation in the UV, IR, spectral, and radio properties of the host of Abell399\_11\_19\_0.

\subsection{SN Classification}

\begin{figure*}
\begin{center}
\includegraphics[width=8cm]{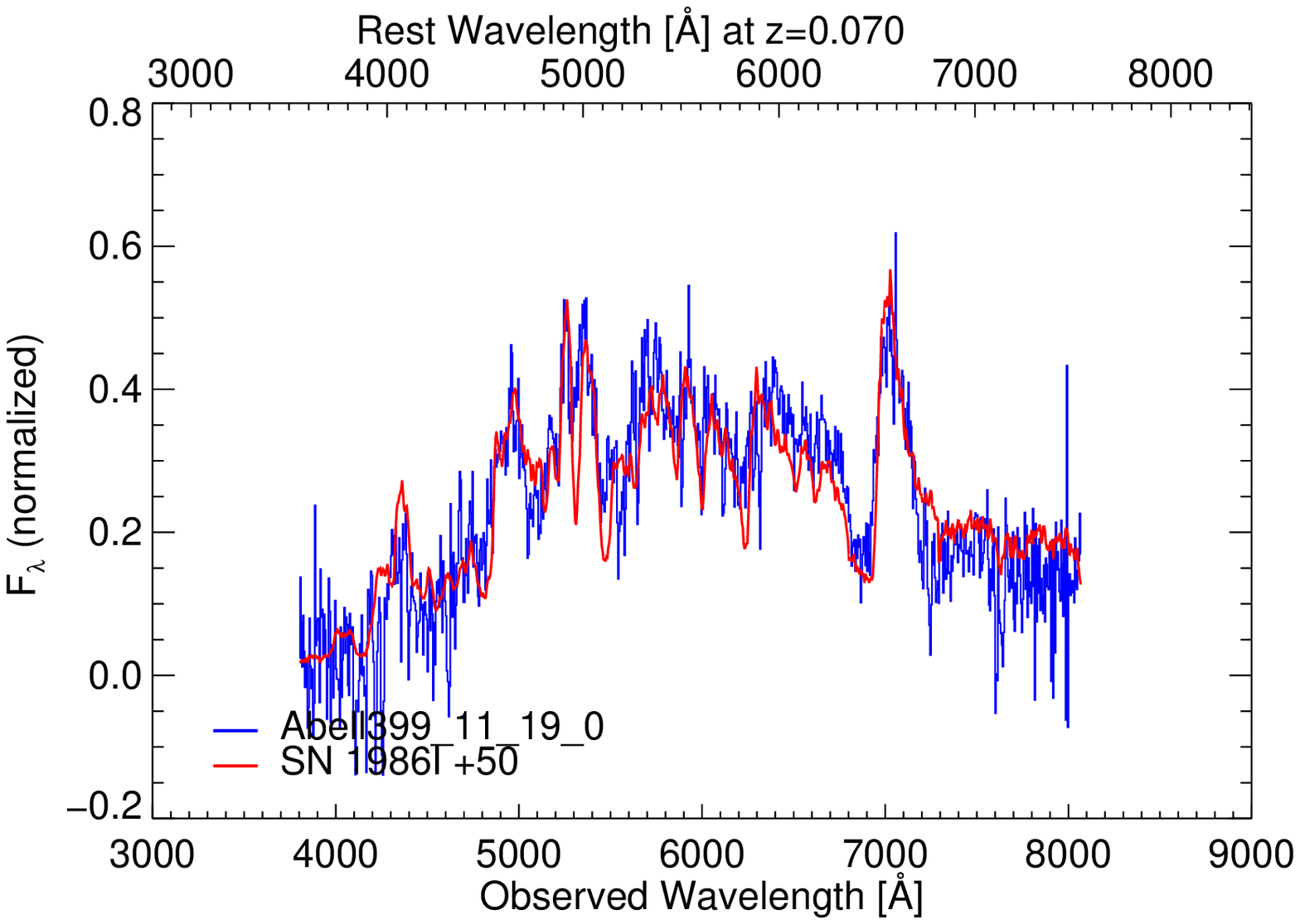}
\includegraphics[width=8cm]{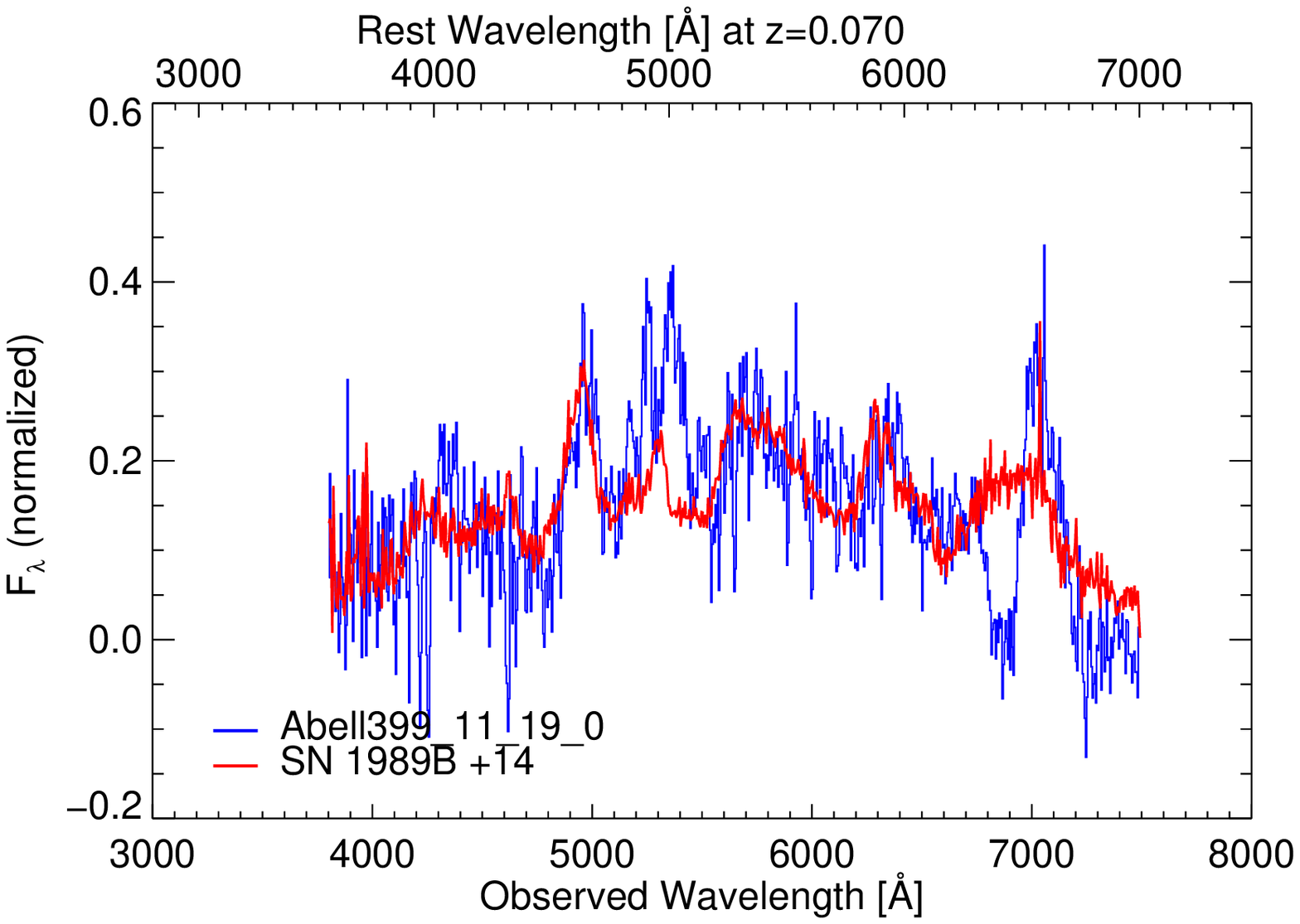}
\caption{The best fitting template from Superfit (red) to the observed spectrum of Abell399\_11\_19\_0 (blue, with best fitting host spectra subtracted), when considering only SN\,II (left) and only SN\,Ia (right) with the spectral template matching code Superfit \citep{Howell05}. The best fit SN\,II template is a much better match than the best SN\,Ia template, in particular to the P-Cygni profile at $\sim7000$ $\rm \AA$. \label{fig:a399_sf}}
\end{center}
\end{figure*}

The classification spectrum taken for Abell399\_11\_19\_0 with Hectospec at the MMT Observatory is shown in Figure \ref{fig:SNset1}, along with the best fitting SN template spectrum from SNID \citep{SNID}. The P-Cygni profile, distinctive for SN\,II, appears at $\sim7000$ $\rm \AA$. Since the fit with SNID may not be overwhelmingly convincing for all our readers, we do an additional analysis with Superfit \citep{Howell05}. This routine achieves a better looking fit because it removes the host galaxy spectrum.  First we run Superfit with loose redshift constraints, $0.05<z<0.25$, to independently confirm the SN redshift. We know the host's spectroscopic redshift is $z=0.072$, and the SN-host association is unambiguous (Figure \ref{fig:a399spec}). The top 3 best fits are SNe\,II at $z=0.07$. We then run Superfit two additional times with the redshift constrained to $z=0.07$: first allowing only SN\,II templates, then SN\,Ia only. The best fits, in Figure \ref{fig:a399_sf}, show that a SN\,II is the better match.

The MENeaCS cadence of one epoch per month did not generate well sampled light curves, but given the relative importance of Abell399\_11\_19\_0 we discuss its photometry briefly.  Following a non-detection epoch in October 2009, this transient was detected in three consecutive months, after which the MENeaCS observations ended. This transient had $m_{g} \sim m_{r} \sim 20$ magnitudes in the detection epoch (corresponding to $M\sim-17.5$ magnitudes). In $\sim55$ days it declined by $\Delta m_{g} \sim 2$ magnitudes and $\Delta m_{r} \sim 1$ magnitude. When compared to the light curve templates of Nugent et al. (2002) and SN luminosity functions of Li et al. (2011a), this color, magnitude, and slow decline are all most consistent with a SN\,II Plateau. Regular SNe\,Ia are too bright, and faint SNe\,Ia decline too quickly (Phillips 1993; Perlmutter et al. 1997). Finally, the preceding non-detection epoch prohibits this from being the late-time shallow-decline epochs of a SN\,Ia.

\subsection{Ultraviolet Photometry}\label{ss:UV}

If this host galaxy experienced a small amount of recent star formation, it might be evident in its near-ultraviolet photometry. Schawinski (2009) generate model spectra from near ultraviolet (NUV) to optical wavelengths by parametrizing the star formation history of early-type galaxies as a large population of old stars plus a small amount of young stars. They show how a galaxy's $NUV-r$ color indicates the time elapsed since the most recent burst of star formation, given the fraction of stellar mass synthesized in the burst. For example, in an early-type galaxy where $\sim1$\% of the stellar mass is $\lesssim50$ Myr old, $NUV-r\lesssim1$. Similarly, Kaviraj (2010) use star formation history models to derive a relationship between the UV and optical photometry of bulge-dominated red galaxies, and the age and mass fraction of their most recent epoch of SF.

We look for the UV counterpart of this host in the GALEX data release 6\footnote{http://galex.stsci.edu/GR6/}. There is no coincident object in the catalog, and under visual inspection the tiles show no hint of a source. From Bianchi et al. (2011), we know that UV sources are detected at 5$\sigma$ down to $NUV\sim20.8$ magnitudes in the GALEX All-sky Imaging Survey, and that the Medium-depth Imaging Survey does not cover this region of sky. With this limit, we restrict the $NUV-r$ color to $>4.5$. Based on Schawinski (2009), this constrains the fraction of stars $<50$ Myr old to $\ll1$\% of the total stellar mass of this SN\,II host galaxy. Comparably, the work presented in Kaviraj (2010) constrains the age and mass fraction of the most recent burst to $\gtrsim 250$ Myr and $<0.5$\%, respectively.

\subsection{Infrared Photometry}\label{ss:IR}

Photometry at optical and near-infrared (NIR) wavelengths can be combined to reveal recent star formation in an evolved galaxy. Stellar population synthesis models have been used by Li et al. (2007) to determine that $B-V$ and $B-K$ photometric colors best disentangle the degeneracy between galaxy age and metallicity. They show that a galaxy's location on a $B-V$ vs. $B-K$ color-color plot can be used to estimate the fraction of mass in young stars. To obtain these colors for this host we begin with our photometry, $m_{g}=17.1\pm0.02$ magnitudes and color $g-r=0.85\pm0.04$, and add 2MASS $m_K=13.62\pm0.18$ magnitudes (from NASA Extragalactic Database). We then K-correct to $z=0$ \citep{Chili10}, and apply filter transformations derived for stars at $z=0$ \citep{Jester05}, to obtain $B-V=0.9\pm0.04$ and $B-K=3.6\pm0.2$. These colors suggest $\lesssim0.5$\% of the stellar mass is younger than 0.5 Gyr \citep{Li07b}.

Galaxies which appear quiescent from optical and NIR photometry can harbor dust-obscured star formation, and be luminous at far-infrared (FIR) wavelengths. In this scenario, dust absorbs UV light and re-emits it in the FIR. Chary \& Elbaz (2001) present conversions from the Spitzer-MIPS 24 $\mu$m wavebands to infrared luminosity, $L_{IR}$, and star formation rates, $SFR_{IR}$. For example, Graham et al. (2010) applied these conversions to Spitzer-MIPS fluxes for 20 optically elliptical SN\,Ia host galaxies from the Supernova Legacy Survey, and found 2 were actually Luminous Infrared Galaxies (LIRGs, $L_{IR}>10^{11}$ $\rm L_{\odot}$) with specific star formation rates $\sim500$ $\rm M_{\odot} \ yr^{-1} \ (10^{12}M_{\odot})^{-1}$.

This red sequence host was observed by the Wide-field Infrared Survey Explorer (WISE; Wright et al. 2010) in its mission to create an all-sky infrared map, and the WISE-W4 filter at 22 $\mu$m compares well with Spitzer MIPS 24 $\mu$m\footnote{Supplement at http://wise2.ipac.caltech.edu/docs/release/prelim}. In W4, this galaxy has an apparent magnitude of 15.2 in the AB system (8.6 in the Vega system), but a signal-to-noise ratio of just 1.4 and a $\sigma=null$, indicating this magnitude is a 95\% confidence upper limit. We convert this magnitude limit to a flux and find $L_{IR}<10^9$ $\rm L_{\odot}$. This is not a LIRG masquerading as a quiescent elliptical.

\subsection{Optical Spectroscopy}\label{ss:spec}

We obtained an optical spectrum of this galaxy with the Blue Channel Spectrograph at the MMT Observatory as part of our program to gather spectra for all our SNe\,Ia cluster hosts. Full spectral analyses will be performed for all of our SN cluster host galaxies in future work (Graham et al. 2012, in preparation).

The partial spectrum presented in Figure \ref{fig:a399spec} reveals hydrogen emission, indicative of star formation -- but also shows the nitrogen, oxygen, and sulfur signatures of a low-ionization nuclear emission-line region (LINER). Kewley et al. (2006) show that star formation is not the dominant source of emission when $\log$$([NII]/H\alpha) > 0.0$ and $\log$$([SII]/H\alpha) > 0.0$. A simple analysis of this galaxy's line intensities finds that [NII] and [SII] are stronger than H$\alpha$ ($\log$$([NII]/H\alpha) \sim 1.1$ and $\log$$([SII]/H\alpha) \sim 0.6$). 
We estimate the maximum amount of H$\alpha$ absorption by fitting template spectra of elliptical galaxies (Kinney et al. (1996); Fioc \& Rocca-Volmerange 1997). After accounting for the template fit with the largest absorption, we are confident that the line intensity of H$\alpha$ is $<4$. This indicates the minimum line ratio values are $\log$$([NII]/H\alpha) > 0.4$ and $\log$$([SII]/H\alpha) \gtrsim 0.0$, which is still consistent with a LINER. While these spectral emission lines mean we cannot attribute the H$\alpha$ to SF, we also cannot rule it out. The slit did cross the galaxy core, and future observations with a slit orientation avoiding the core may reveal SF at this SN's location.

\subsection{Radio Power}\label{ss:radio}

Radio-loud emission ($\rm L_{1.4GHz}>10^{29} \ erg \ s^{-1} \ Hz^{-1}$) from elliptical galaxies could indirectly represent ongoing star formation (e.g. Della Valle et al. 2005). We checked published radio source catalogs for 1.4 GHz emission at this galaxy's position. Its coordinates are not covered by the VLA FIRST Survey\footnote{http://sundog.stsci.edu/}, and it was not detected in the NRAO VLA Sky Survey\footnote{http://www.cv.nrao.edu/nvss/} (NVSS). The completeness limit of the NVSS at 1.4 GHz is 2.5 mJy. At the redshift of this host, $z=0.072$, the radio-loud population is incomplete and we cannot constrain the radio properties of this host.

\smallskip

In summary we find no evidence of star formation in this galaxy, aside from the presence of the SN\,II. This indicates that either very low levels of star formation and trace amounts of young stellar populations can exist in red sequence galaxies, or there is a rare other channel to SNe\,II with a longer delay time.

\section{The Cluster SN\,II Rate} \label{s:rate}

Our calculation of the SN\,II rate in clusters follows the method we used for SN\,Ia rates in S12, which is very similar to that used for high and low redshift SN rates by Sharon et al. (2007) and Barbary et al. (2012). The rate of Type II supernova, $SNR_{II}$, is calculated by:

\begin{equation} \label{e:rate}
SNR_{II} = \frac{N_{II} \times C_{inc}/C_{spec}}{\sum_{j=1}^{j=N_{ep}}\Delta t_j M_j} ,
\end{equation}

\noindent
where $N_{II}$ is the observed number of SNe\,II. The spectroscopic completeness,$C_{spec}=0.91$, accounts for the $\sim9$\% of the time when MENeaCS was detecting SNe, but we did not have spectroscopic follow-up due to weather and telescope scheduling. This value is independent of SN type (see S11 and S12). The inclination correction, $C_{inc}$, accounts for SNe\,II that are undetectable due to extreme dust obscuration in highly inclined and edge-on spiral galaxies (e.g. Cappellaro et al. 1993b; Cappellaro et al. 1999); our inclination correction, $C_{inc}=1.62$, is derived in Appendix A. Over all $N_{ep}$ survey epochs of every cluster we sum the control time for that epoch, $\Delta t_j$, multiplied by the mass or luminosity surveyed in that epoch, $M_j$ (see S12 for a description of how these are calculated from our deep image stacks). 

The control time is the effective amount of time surveyed by the $j^{th}$ epoch, expressed by:

\begin{equation} \label{e:tctrl}
\Delta t = \int_{t_1}^{t_2}  \eta (m(t)) dt, 
\end{equation}

\noindent
where $\eta(m)$, the MENeaCS detection efficiency as a function of apparent magnitude, is determined from simulated transients and is presented in S12. Although the population of simulated transients used for our recovery statistics have magnitude distributions that mimic a sample of SNe\,Ia, the resulting detection efficiency is appropriate for use with our SNe\,II. For the SN\,II light curve, $m(t)$, we start with the absolute $V$-band SN\,II template light curves from Nugent et al. (2002). MENeaCS detection efficiencies were calculated in the $g$-band. We convert from $M_V$ to $m_{g}$ using the cluster's redshift, the SN\,II K-correction (based on spectral templates from Nugent et al. 2002), and the photometric calibrations for the $j^{th}$ epoch. The integration boundaries $t_1$ and $t_2$ are defined by the time during which the SN\,II template light curve meets our color limit for spectroscopic follow-up, $g-r < 0.8$ (e.g. 39 days for $z=0.15$, and 66 days for $z=0.05$). We account for potential re-discoveries of the same transient in multiple epochs by subtracting from $\eta(m)$ the probability that it was detected previously, in the same fashion as S12 and Sharon et al. (2007).

We use a Monte Carlo method in which the rate is calculated many times. For each realization a peak absolute magnitude is randomly chosen from the luminosity function discussed in \S~\ref{ss:LF}. At every instance of $M_j$, $\eta(m)$, and $N_{II}$, we randomly draw their value from an appropriate distribution based on their uncertainty (e.g. Poisson error for $N_{II}$). We run this Monte Carlo calculation for 500 realizations, which generates a distribution of rates. The final value for the rate is the median of this distribution, and the statistical uncertainties correspond to the 16th and 84th percentiles (the 68\% confidence interval). The final results are presented in \S~\ref{ss:results}.

\subsection{The SN\,II Luminosity Function} \label{ss:LF}

For each realization of the Monte Carlo we randomly draw the SN\,II subtype (P, L, b, or n) and peak absolute magnitude from the volume-limited luminosity functions (LFs) published by the LOSS (Li et al. 2011a; hereafter Li11a). We use their LFs for type S0-Sbc hosts because the majority of the stellar mass in cluster environments is in early-type galaxies. The fractions of each SN\,II subtype in S0bc host galaxies are: P, 72\%; L, 12\%; b, 9\%; and n, 7\%. These fractions do not change by more than 2\% when all host types are considered, and the SN\,II LF for all host types is very similar to that for S0bc only. We also use the appropriate light curve template for each subtype from Nugent et al. (2002); the SN\,IIL light curve when subtype L is chosen, and the SN\,IIP light curve when subtypes P, b, or n are chosen. 

\begin{figure*}
\begin{center}
\includegraphics[width=8cm]{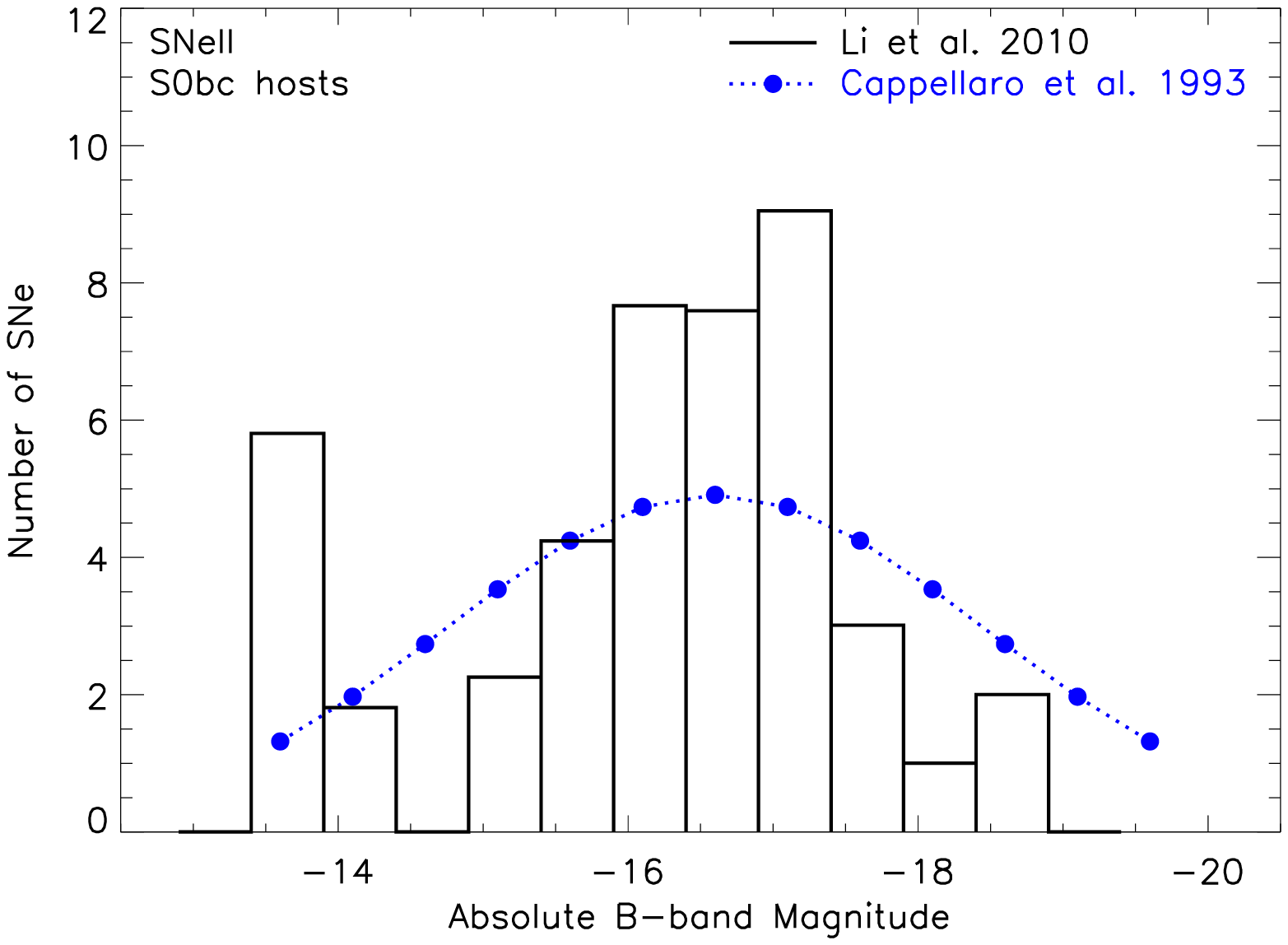}
\includegraphics[width=8cm]{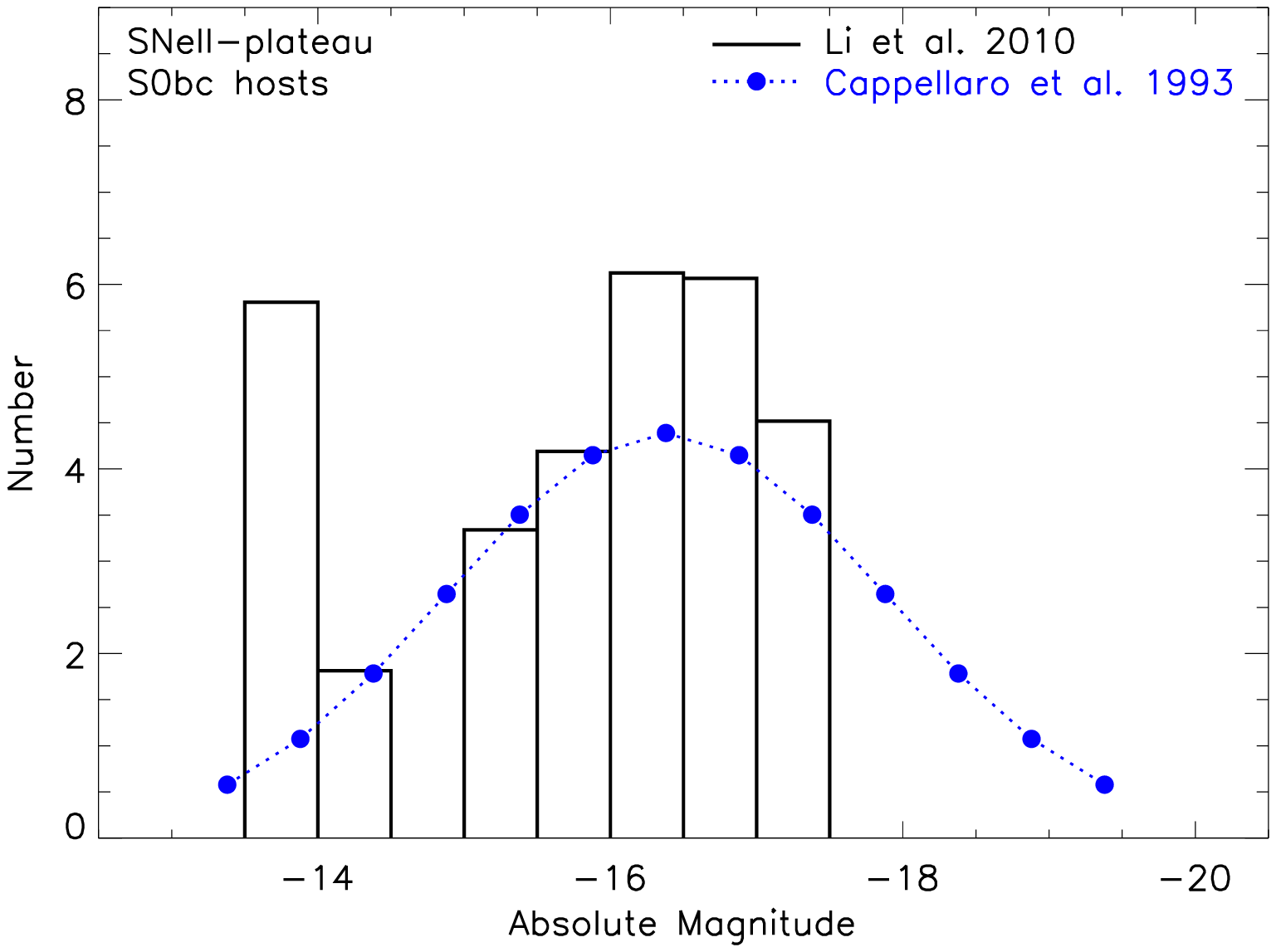}
\caption{SN\,II luminosity functions. Black, the volume-limited SN\,II luminosity function observed by Li et al. (2010), for host galaxies of type S0-Sbc. Blue, the SN\,II luminosity function used by Cappellaro et al. (1993a), normalized to the same total number as the Li et al. (2010). \label{fig:SNLF} }
\end{center}
\end{figure*}

To compare to the $SNR_{II}$ from M08, we repeat the Monte Carlo process using the same SN\,II LF as them: that of Cappellaro et al. (1993a; hereafter C93a). They present the SN\,II LF as Gaussian functions for subtypes P and L separately: peak $M_{B,IIP}=-16.38$, $\sigma_{IIP}=1.49$, peak $M_{B,IIL}=-16.82$, and $\sigma_{IIL}=1.1$ magnitudes. In Figure \ref{fig:SNLF} we compare the SN\,II LFs from Li11a and C93a. To plot a single LF from C93a, we combine the IIP and IIL into one Gaussian of peak $M_{B}=-16.51$ magnitudes, $\sigma=1.85$ magnitudes, assuming 30\% SN\,IIL and 70\% SN\,IIP. As is evident in Figure \ref{fig:SNLF}, Li11a detects a population of faint SN\,II-P ($M_B\sim-14$ magnitudes), which results in a non-Gaussian LF. However, since MENeaCS is not sensitive to transients fainter than $M_B\sim-14.6$ magnitudes, these two LFs produce effectively similar results. This is discussed further in \S~\ref{ss:caveats}.

Here we make two important notes about how we incorporate the SN\,II LFs and light curves. First, the Li11a distribution of absolute peak magnitudes from their unfiltered survey is very closely matched to R-band, and can be considered as $M_R$ magnitudes with no correction. The C93a LF is for $M_B$, which peaks several days earlier, and the Nugent et al. (2002) light curve templates are for $M_V$. Fortunately, we do not need to convert between filters because the intrinsic $B-V$ and $V-R$ colors of SNe\,II at the time of $B$- and $V$-band maximum light is $\sim0$ (e.g. Poznanski et al. 2002; d'Andrea et al. 2010). Although SNe\,II will be redder a few days later, at the time of maximum light in the $R$-band this is due to a decline in $B$ and $V$; the $R$-band magnitude increases only slightly between the times of $B$- and $R$-band maximum light. Therefore, we directly apply the $R$-band LF of Li11a and the $B$-band LF of C93a to the $V$-band light curve templates of Nugent et al. (2002).

Second, neither Li11a nor C93a correct their SN\,II LFs for host extinction, and by choosing randomly from these LFs we automatically include host extinction in our Monte Carlo rate calculation. By using the Li11a LF for S0-Sbc type hosts, the host extinctions are as similar to that expected for cluster galaxies as possible because most cluster galaxies are of similar early types. While the observed colors of SNe\,II-P do have a spread due to host reddening (e.g. Hamuy 2003; Krisciunas et al. 2009; Olivares E. et al. 2010), this affect is expected to be small for most of our surveyed mass in galaxy clusters. For example, Hamuy (2003) find that SNe\,II associated with groups/clusters show little to no reddening. The affect of host reddening and extinction on $SNR_{II}$ is accounted for by the combination of our chosen LFs because they are uncorrected for host dust, and by our inclination correction which is discussed below.

\subsection{Interlopers}
\label{ss:interlopers}

As mentioned in \S~\ref{s:SNII}, cluster membership was assigned for SNe with $|v_{SN}-v_{cluster}|<3000$ $\rm km \ s^{-1} $, which actually includes $\sim50$ Mpc in front of and behind each cluster. Any SNe\,II exploding in Hubble flow galaxies within this cylindrical volume may be erroneously associated with our galaxy clusters. The number of interlopers we expect to have observed, $N_{exp}$, after accounting for our detection efficiencies is:

\begin{equation} \label{e:nexp}
N_{exp} = \frac{R_{vol} \ C_{spec}}{C_{inc}} \ \sum_{j=1}^{j=N_{ep}} \Delta t_j V_j .
\end{equation}

The volumetric SN\,II rate at $z\sim0.1$ is $R_{vol} \sim 7\pm3 \times 10^{-5}$ $\rm SN \ yr^{-1} \ Mpc^{-3}$ \citep{bazin09}. The MENeaCS spectroscopic completeness term $C_{spec}=0.91$. The inclination correction, $C_{inc}$, is discussed in Appendix A. We assume an interloper-hosting field galaxy would not be elliptical, which slightly raises the inclination correction factor used here to $C_{inc}=1.72$. The final term is the control time, $\Delta t$, multiplied by $V_j$, the comoving volume element within $\pm3000$ $\rm km \ s^{-1}$ and the chosen cluster radius (1 Mpc or $R_{200}$). This term is summed over all observed epochs, $\rm N_{ep}$. Our control times are shorter than the 1 year of MENeaCS survey time because SNe\,II are intrinsically fainter than the SNe\,Ia which MENeaCS was designed to find. 

For a cluster radius of 1 Mpc, the result is an expected number of interloping supernovae $N_{exp}=0.2_{-0.1}^{+0.2}$. The Poisson probability that we observed 0, 1, or 2 SNe within 1 Mpc is $\sim0.81$, 0.17, and 0.02 respectively. Similarly for a cluster radius of $R_{200}$, $N_{exp}=0.7_{-0.3}^{+0.5}$, and the probabilities of observing 0, 1, 2, or 3 SNe are $\sim0.50$, 0.35, 0.12, and 0.03. For every realization of our Monte Carlo we randomly draw a value of $N_{exp}=0$, 1, 2, or 3 SNe, weighted by its respective probability, and subtract it from the number observed in order to produce SN\,II rates statistically corrected for interloping SNe. Since our one red sequence SN\,II host has a redshift consistent with the cluster {\it and} lies on the photometric red sequence -- and considering that the number density of potential red sequence interlopers is relatively small -- we consider it very unlikely to be an interloper, and we do not apply the interloper correction to the red sequence $SNR_{II}$.

\subsection{MENeaCS SN\,II Rates and Uncertainties}\label{ss:results}

The MENeaCS SN\,II cluster rates are presented in Table \ref{table:SNrates} in the conventional units of SNuB and SNuM, where SNuB $\equiv$ SNe(100 yr $10^{10} L_{B,\odot}$)$^{-1}$ and SNuM $\equiv$ SNe(100 yr $10^{10} M_{\odot}$)$^{-1}$. We present the rate in three cluster galaxy subsets: ``All", the total stellar mass including the intracluster stars; ``RS", red sequence galaxies only; and ``Off RS", galaxies lying off their cluster's red sequence. We show our results with the SN\,II luminosity functions from both Li11a and C93a, and within cluster radii of 1 Mpc and $R_{200}$.

The statistical uncertainties include the Poisson error on the number of SNe\,II observed, the interloper contamination, the uncertainty on cluster luminosity, and the uncertainty in our detection efficiencies. Statistical errors are dominated by the Poisson uncertainties and interloper contamination (with relative contributions of $\sim$2/3 and $\sim$1/3, respectively), with very small uncertainties (1$\sim$5\%) from our detection efficiency and cluster mass/luminosity uncertainty. The systematic uncertainties include a small contribution from the inclination correction factor (5\%; Appendix A), an offset of $\sim10$\% in cluster luminosity which was derived from a comparison of MENeaCS photometry to SDSS, and the uncertainty in $R_{200}$ for rates within $R_{200}$ (the latter two are discussed in detail in S12). The relative contributions to the systematic uncertainty from these two components is $\sim30$\% and 70\% respectively.

\subsection{Caveats}\label{ss:caveats}

Here we make several important notes regarding the MENeaCS SN\,II rate in clusters. First, ZwCl0628\_7\_08\_0 was both $m_{g}>22.5$ magnitudes and $g-r>0.8$ at all detections, and therefore did not officially meet our follow-up requirements. Occasionally we followed-up targets outside our formal bounds, but despite its spectroscopic confirmation we cannot include ZwCl0628\_7\_08\_0 in the rates.

Second, our control time goes to zero for SN\,II fainter than $M_B\sim-14.6$ magnitudes, yielding an unphysical infinite rate. We exclude these realizations when calculating the median rate and its uncertainty, which is effectively the same as truncating the SN\,II LF at $M\sim-14.6$ magnitudes. This approach is valid because it is inappropriate to calculate the rates of objects to which a survey is insensitive. The result is that our rates are for the ``normal" population of SNe\,II with $M_B\leq-14.6$ magnitudes, and do not include the faint sub-population found by Li11a. If we include the realizations of infinite rates in the median (e.g. for ``All" galaxy types and $R_{200}$), the rate becomes 0.031 instead of 0.026 SNuM with the Li11a LF -- a difference of $\lesssim0.07\sigma$. For the C93a LF, the difference is also negligible. In Appendix B we discuss a rate calculation method which integrates over the SN\,II LF to avoid the instances of zero control times, and explain how it is not appropriate for MENeaCS.

Third, as discussed in \S~\ref{s:SNII} our sparse light curve sampling and single epoch spectroscopy means we cannot identify SN\,II subtypes P, L, b, or n. It is likely that the subtype distribution in clusters is similar to that in field S0-Sbc hosts reported by Li11a, so it is possible that our sample contains all SN\,IIP. For this reason, we also run our Monte Carlo using SN\,II LFs for the plateau subtype only. Figure \ref{fig:SNLF} shows that the SN\,IIP LFs are fainter, and that the L, b, and n subtypes populate the magnitude bins brighter than -17.5. This leads to higher rates for SN\,IIP: 0.032 instead of 0.026 SNuM with the Li11a LF, which is a difference of $\lesssim0.08\sigma$.

\subsection{Comparison to Published SN\,II Rates} \label{ss:comp}

The only previous measure of the cluster $SNR_{II}$ is from M08, who compiled five visual and photographic $z<0.02$ galaxy-targeted SN searches in which ``the original galaxy sample was not selected in order to reproduce the cosmic average but rather to have a significant number of SN detections". This means that massive galaxies were preferentially targeted and associated with galaxy clusters later, and that the surveyed cluster mass of M08 is incomplete. Furthermore, SNe discovered during the surveys used by M08 were not all spectroscopically classified. MENeaCS, on the other hand, surveyed galaxy clusters between $0.05<z<0.15$ to $R_{200}$, including the stellar mass in faint galaxies and the intracluster population, and uses only spectroscopically confirmed SNe. Although the MENeaCS survey strategy is better, our survey duration is shorter, our discovered number of SNe\,II is smaller, and the Poisson uncertainties are larger. In general the M08 rates in Table \ref{table:IIrates} and the MENeaCS rates in Table \ref{table:SNrates}, for the variety of cluster radii and galaxy types considered, are consistent at the 1--2$\sigma$ level.

A comparison of the rate of core collapse supernovae ($SNR_{CC}$, which includes Types II and Ibc) between cluster and field galaxies with similar SFRs can potentially reveal an environmental dependence of the initial mass function (IMF). For example, consider two galaxies with the same total star formation rate: if one has a higher $SNR_{CC}$, then it is forming a larger fraction of $\geq8$ $\rm M_{\odot}$ stars. We did not discover any SN\,Ibc. The Ibc:II ratio is typically 1:2 to 1:4 in cluster and field environments, respectively (M08; Li et al. 2011b). Based on this, we only expect 1.7--3.5 SNe\,Ibc, and the probability of observing zero is 3--18\%. Thus, we are insensitive to any environmental dependence of the IMF at masses $\geq20$ $\rm M_{\odot}$. A comparison of the MENeaCS $SNR_{II}$ in cluster ``RS" and ``Off-RS" galaxies to the published rates in field E/S0 and S0a/b galaxies in Table \ref{table:IIrates} finds a general 1--2$\sigma$ agreement. We conclude that SN\,II cluster rates require a more precise measurement in order to firmly identify any difference in the cluster IMF.

\section{Cluster Star Formation Rates} \label{s:analysis}

We begin our discussion with the first-ever derivation of cluster SFR from SNe\,II, which we compare with other measurements of cluster SFR, in \S~\ref{ss:sfr}. In \S~\ref{ss:icsfr} we discuss the MENeaCS limits on intracluster star formation. In \S~\ref{ss:AB} we discuss the implications of cluster SF for SNe\,Ia, including whether short-delay SNe\,Ia contaminate the late-time DTD when measured via the cluster SN\,Ia rate; the possibility that all SNe\,Ia have short delays; and the potential source of the $SNR_{Ia}$ enhancement in cluster ellipticals observed by M08. 

\subsection{Derivation of Cluster SFR from SN\,II Rates} \label{ss:sfr}

Botticella et al. (2012) present a derivation of the relation between the CC\,SN rate and $SFR$, which we adapt to SN\,II: 

\begin{equation} \label{e:sfr}
R_{II} = K_{II} \times SFR .
\end{equation}

\noindent
In the above expression, $R_{II}$ is the rate of SN\,II with units of $\rm SN \ yr^{-1}$ (whereas $SNR_{II}$ is in units of SNuM = $\rm SN \ (10^2 yr)^{-1} \ (10^{10} M_{\odot})^{-1}$). The value of $R_{II}$ is calculated in a similar manner as $SNR_{II}$, and presented in Table \ref{table:SFR}. The $SFR$ is in units of $\rm M_{\odot} \ yr^{-1}$, and $K_{II}$ is the number fraction of all stars formed which explode as SN\,II ($\rm SN \ M_{\odot}^{-1}$): 

\begin{equation}
K_{II} = \frac{\int_{m_{l,II}}^{m_{u,II}} \phi(m) dm}{\int_{m_l}^{m_u} m \phi(m) dm}.
\end{equation}

\noindent
The integration limits $m_{l,II}$ and $m_{u,II}$ correspond to the minimum and maximum initial masses of stars which become SNe\,II. The lower limit is generally agreed to be $\sim8$ $\rm M_{\odot}$, and evidence is converging towards an upper limit of $\lesssim20$ $\rm M_{\odot}$ (e.g. Smartt et al. 2009; Dessart et al. 2010). For $\phi(m)$, the initial mass function (IMF), we use the Salpeter expression where $\phi(m) \propto m^\gamma$, and $\gamma=-2.35$ (Salpeter 1955). The limits $m_{l}$ and $m_{u}$ are the mass range of the IMF, for which we use 0.1 and 100 $\rm M_{\odot}$. Under these assumptions, $K_{II}=0.0054$. By extending $m_{l}$ and $m_{u}$ to 0.05 and 200, testing a slightly lower exponent of $\gamma=-2.3$ (Kroupa 2001), and considering a lower $m_{u,II}=17$ $\rm M_{\odot}$, we estimate an uncertainty on $K_{II}$ of $\pm0.001$. We account for this with a $\sim18$\% systematic on the cluster $SFR$. We also calculate the specific star formation rate as $sSFR = SNR_{II} / K_{II}$, and list them in Table \ref{table:SFR}.

We now compare with a selection of previous cluster SFR measurements from IR and UV imaging, and optical spectroscopy. Chung et al. (2011) used WISE images of 72 low redshift ($z<0.1$) galaxy clusters at wavelengths 3.4, 4.6, 12, and 22 $\mu$m. They found that for $0.5<R<1 \ R_{200}$, clusters have a mean $sSFR \sim 2$ $\rm M_{\odot} \ yr^{-1} \ (10^{12}M{\odot})^{-1}$ (see their Figure 3). Our measurement for ``All" galaxy types in clusters, $sSFR = 5.1_{-3.1}^{+15.8}\pm0.9$ $\rm M_{\odot} \ yr^{-1} \ (10^{12}M{\odot})^{-1}$, is in agreement with this. Yi et al. (2005) used GALEX UV images to look for evidence of recent SF in SDSS field early-type galaxies. They found that 1--2\% of the stellar mass in $\sim15$\% of bright ($M_R<-22$ magnitudes) early-types formed within the last $\sim1$ Gyr. If this is true for clusters, we would expect $SFR=2$--4 $\rm M_{\odot} \ yr^{-1}$ in bright red sequence galaxies. If true for fainter red sequence members also, the expected total SFR raises to 10--20 $\rm M_{\odot} \ yr^{-1}$, which is $\sim5$ times higher than (but within 2$\sigma$ of) our red sequence SFR, $2.1_{-0.9}^{4.2}\pm0.4$ $\rm M_{\odot} \ yr^{-1} \ (10^{12}M_{\odot})^{-1}$. Based on the H$\alpha$ luminosity from SDSS optical spectra, Finn et al. (2008) determined that $sSFR\sim9$ $\rm M_{\odot} \ yr^{-1} \ (10^{12}M_{\odot})^{-1}$ for star-forming galaxies in low redshift clusters. This is in 1$\sigma$ agreement with our ``Off-RS" $sSFR=16.6_{-9.6}^{+39.5}\pm3.0$ $\rm M_{\odot} \ yr^{-1} \ (10^{12}M_{\odot})^{-1}$. 

Finally, we note that our low $sSFR$ in red sequence galaxies implies that only $\sim0.01$\% of the mass in the red sequence at $z\sim0.1$ is comprised of stars formed in the past 50 Myr. This extremely low percentage is consistent with the lack of detection in UV and IR of the red sequence host of Abell399\_11\_19\_0 presented in \S~\ref{s:RShost}.

\subsection{Intracluster Star Formation} \label{ss:icsfr}

Direct evidence of intracluster star formation has been presented by Sun et al. (2010), who detected a 40 kpc long X-ray tail extending from a galaxy in a nearby rich cluster. With optical spectra they identified 35 H II regions along this tail, the furthest of which are 20 kpc away from the galaxy (far enough to be defined as intracluster). Also, simulations of galaxy clusters investigating the size and origin of the IC stellar population suggest that $\sim30$\% of the IC stars form at significant distances from a galaxy dark matter halo \citep{puchwein10}. In such simulations most of the IC stars form at $z>1$, with just a small tail of $\sim1.5$\% of the final IC stellar mass forming during the last $\sim1.8$ Gyr (since $z=0.15$). For the average IC stellar mass of our clusters, assuming an IC mass fraction of 16--45\% (Gonzales et al. 2005; S11), this implies an IC SFR of 5--13 $\rm M_{\odot} \ yr^{-1}$, and $R_{II}=0.019$--0.049 SN\,II $\rm yr^{-1}$. This converts into an expectation of 0.3--0.8 IC SNe\,II in the MENeaCS sample, which is consistent with our observation of zero. Interestingly, if the Puchwein et al. (2010) simulations are correct, the first detection of an IC SN\,II would be likely in a survey just twice as large as MENeaCS.

Although we know our upper limit on the IC SFR will not be very restrictive, it is the first derived from the non-detection of IC SNe\,II. This is only possible in a complete, well characterized survey like MENeaCS. To calculate an upper limit for the rate of IC SNe\,II within $R_{200}$, we use Poisson statistics for a detection of zero \citep{gehrels86}, and allow fractional values of IC $N_{II}$ in the Monte Carlo rate calculation represented by Equation \ref{e:rate} (with $C_{inc}=1.0$). The resulting 1$\sigma$ upper limit is $R_{II}<0.15$ $\rm SN \ yr^{-1}$, which converts to an upper limit of IC $SFR < 28$ $\rm M_{\odot} \ yr^{-1}$.

\subsection{Implications For Cluster SNe\,Ia} \label{ss:AB}

In \S~\ref{s:intro} we described how deducing the late-time SN\,Ia DTD from the cluster $SNR_{Ia}$ as a function of redshift must assume that cluster stars all formed in a burst at high redshift, and that all cluster SNe\,Ia have experienced a long, $>2$ Gyr, delay time. In this section we explore the impact of cluster star formation on this assumption, and its implications for the SN\,Ia DTD. First, we discuss our results with respect to the possibility of prompt-only SN\,Ia DTD. Next, we estimate the contamination of ``prompt" SNe\,Ia to the cluster sample at low redshift. Finally, we comment on the implications of our cluster $SNR_{II}$ and $SFR$ regarding the source of the enhanced $SNR_{Ia}$ in cluster ellipticals presented by M08.

\subsubsection{A DTD of Short Delays Only?}

Maoz et al. (2010) find that the cluster $SNR_{Ia}$ as a function of redshift is best and most simply fit by a brief burst of star formation at $z\sim3$ combined with a SN\,Ia DTD that peaks at short delays of $<2$ Gyr and decreases as a power law with a slope of -1 to long delays of $\sim11$ Gyr. This DTD is consistent with theoretical predictions of the double degenerate scenario. They also consider a DTD of short delays only, congruent with predictions for the single-degenerate model. They find it can only reproduce the observed cluster $SNR_{Ia}(z)$ from $0<z<1.2$ if they include ongoing cluster star formation within a radius of 1 Mpc.

Maoz et al. (2010) reject the prompt-only DTD hypothesis in part because previous surveys found SNe\,Ia mainly within 1 Mpc and always in elliptical, red sequence galaxies showing no signs of star formation, and in part due to other work showing cluster star formation was predominantly outside of 1 Mpc. With MENeaCS, we have shown that SNe\,Ia do occur outside of 1 Mpc, and in blue cluster galaxies (Figure \ref{fig:RoffCC}, and S12). We have also shown that SNe\,II, and therefore star formation, occurs inside of 1 Mpc and in red sequence galaxies. 

Qualitatively, this suggests it would be premature to rule out a DTD of short delays only, and the single-degenerate model as the sole scenario, based on observed SNe\,Ia cluster rates. Quantitatively, Maoz et al. (2010) find that a SN\,Ia DTD of short delays only requires a rate of ongoing star formation $SFR\sim175$ $\rm M_{\odot} \ yr^{-1}$ in the central regions of galaxy clusters. This is $\sim3$ times higher than our cluster SFR inferred from SNe\,II, but our uncertainties on cluster SFR are large, and $175$ $\rm M_{\odot} \ yr^{-1}$ is actually just within the 1$\sigma$ upper limit. Ultimately, although our cluster SFR rate is low, we cannot rule out a DTD of short delays only.

\subsubsection{Fraction of ``Prompt" Cluster SNe\,Ia}

The SN\,Ia rate per unit mass, $SNR_{Ia}$, is a convolution of star formation history of the surveyed galaxy sample and the SN\,Ia DTD. As an oversimplified parametrization, it can be expressed as the sum of ``delayed" and ``prompt" components represented by constants $A$ and $B$, where $SNR_{Ia} = A + B \times sSFR$ \citep{mann05,sb05,Sullivan06}. Based on this, the fraction of SNe\,Ia expected to be associated with the prompt component is: $f_{prompt} = (B \times sSFR)/SNR_{Ia}$. We use the $B$ value from Sullivan et al. (2006), $3.9 \pm 0.7 \times 10^{-4}$ $\rm SN \ yr^{-1} \ (M_{\odot}yr^{-1})^{-1}$. For MENeaCS red sequence cluster galaxies, our derived sSFR is in Table \ref{table:SFR} and $SNR_{Ia} = 0.041_{-0.015}^{+0.015}$$_{-0.010}^{+0.005}$ SNuM (S12). This reveals $f_{prompt} = 0.02_{-0.02}^{+0.05}$ in cluster red sequence galaxies; similarly, $f_{prompt} = 0.05_{-0.05}^{+0.19}$ for ``All" cluster galaxies. 

While this minimal $\sim$2\% contamination indicates that all of the MENeaCS SNe\,Ia likely experienced a long delay time, we cannot rule out that up to 7\% of SNe\,Ia in red sequence hosts exploded with short delay times (the 1$\sigma$ confidence level). If so, then the ``RS" $SNR_{Ia}$ for the delayed SNe\,Ia only would be $\sim0.038$ instead of $0.041$ SNuM. This is a difference of 0.003 SNuM or, expressed in terms of the uncertainty on $SNR_{Ia}$,  $\sim0.2$$\sigma$. Ultimately, we find the potential maximum contamination from ``prompt" SNe\,Ia is less than our statistical uncertainty on the cluster SN\,Ia rate. However, this may not be the case for higher redshift SN\,Ia cluster surveys.

\subsubsection{The Enhanced $SNR_{Ia}$ in Cluster Ellipticals}

The rate of SNe\,Ia in cluster early-type galaxies was found to be a factor of three higher than the rate in field early-types by M08: $SNR_{Ia}=$ 0.066 SNuM compared to 0.019 SNuM, respectively. They report that if this excess is from the ``prompt" component, they would only expect $\sim2$ SNe\,II -- consistent with their detection of no SNe\,II in E/S0 hosts.  In S12, we report the cluster red sequence $SNR_{Ia} = 0.041_{-0.015}^{+0.015}$$_{-0.010}^{+0.005}$ SNuM, which is an excess of 0.022 SNuM over the rate in field early-types from M08 (but given the uncertainties still consistent). Assuming this is entirely from the prompt component, and using the $B$ value from Sullivan et al. (2006), we find an implied $sSFR = 56_{-45}^{+38}$ $\rm M_{\odot} \ yr^{-1} \ (10^{12}M_{\odot})^{-1}$
in cluster red sequence galaxies. This is a factor of $\sim28$ times greater than our observed SFR in Table \ref{table:SFR}. If we instead consider our ``All" galaxy sample the implied $sSFR = 56_{-34}^{+40}$ $\rm M_{\odot} \ yr^{-1} \ (10^{12}M_{\odot})^{-1}$ is a factor of $\sim11$ times greater than the observed SFR in clusters. Given that the relation between SFR and $SNR_{II}$ is direct (Equation \ref{e:sfr}), then we should have observed an order of magnitude more SNe\,II in MENeaCS. Our data suggests that either the SN\,Ia rate enhancement in cluster ellipticals does not exist, or it is not due to recent star formation.

\section{Conclusion}\label{s:conc}

In this paper we present the 7 SNe\,II discovered in $0.05<z<0.15$ rich galaxy clusters by MENeaCS. Our sample also includes one SN\,II in a red sequence galaxy which shows no clear evidence of recent star formation in its multi-wavelength properties. This illustrates the danger of using host morphology to classify SNe in lieu of expensive spectroscopy time. The simplest explanation is that undetectable levels of star formation exist in the elliptical host. If that is not the case, it leaves open the possibility of a rare other channel to SNe\,II with a long delay time.

With the MENeaCS sample we make the first measurement of $SNR_{II}$ from a survey which is both cluster-targeted and complete to $R_{200}$. We also make the first derivation of cluster SFR from $SNR_{II}$, and find that it agrees with SFR measurements for cluster galaxies and field ellipticals from IR and UV photometry, and H$\alpha$ line emission. We show how these low levels of cluster star formation imply that a small fraction of cluster SNe\,Ia may have experienced a short delay time. However, we find their influence on the cluster $SNR_{Ia}$ is within statistical uncertainties, and does not undermine the use of low redshift cluster $SNR_{Ia}(z)$ to derive the delay time distribution for SNe\,Ia.

\acknowledgments
 
We gratefully acknowledge the CFHT Queued Service Observations team, without whom MENeaCS would not have been possible. 
We thank Nelson Caldwell for managing the MMT/Hectospec queue, and also thank Stephenson Yang for his dedication to essential computer and network maintenance.
CJP acknowledges support from the National Engineering and Science Research Council of Canada.
HH acknowledges support from a Marie Curie International Reintegration Grant and the NWO Vidi grant.
This work is based in part on data products produced at the Canadian Astronomy Data Centre as part of the Canada-France-Hawaii Telescope Legacy Survey, a collaborative project of NRC and CNRS. 
This work is based on observations obtained with MegaPrime/MegaCam, a joint project of CFHT and CEA/DAPINA, at the Canada-France-Hawaii Telescope (CFHT) which is operated by the National Research Council (NRC) of Canada, the Institut National des Sciences l'Universe of the Centre National de la Recherche Scientifique (CNRS) of France, and the University of Hawaii.
Observations reported here were obtained at the MMT Observatory, a joint facility of the Smithsonian Institution and the University of Arizona. 
This research has made use of the VizieR catalogue access tool, CDS, Strasbourg, France. 
This research has made use of the NASA/IPAC Extragalactic Database (NED) which is operated by the Jet Propulsion Laboratory, California Institute of Technology, under contract with the National Aeronautics and Space Administration. 
This research has made use of the NASA/IPAC Infrared Science Archive, which is operated by the Jet Propulsion Laboratory, California Institute of Technology, under contract with the National Aeronautics and Space Administration.
This publication makes use of data products from the Wide-field Infrared Survey Explorer, which is a joint project of the University of California, Los Angeles, and the Jet Propulsion Laboratory/California Institute of Technology, funded by the National Aeronautics and Space Administration.

{\it Facilities:} \facility{CFHT, MMTO, Gemini}.

\clearpage

\bibliographystyle{apj}
\bibliography{apj-jour,mybib}

\clearpage

\begin{deluxetable*}{llcc}[h]
\tablecolumns{4}
\tablecaption{Published SN\,II Rates \label{table:IIrates}}
\tablehead{
\colhead{Environment} & \colhead{Galaxy} & \colhead{Rate} & \colhead{Rate} \\\colhead{} & \colhead{Types} & \colhead{SNuB\tablenotemark{a}} & \colhead{SNuM\tablenotemark{b}} 
}
\startdata
\multicolumn{4}{l}{Mannucci et al. (2008)} \\
\hline
Cluster & Total         &  0.23$^{+0.09}_{-0.07}$   &  $0.072^{+0.028}_{-0.021}$   \\
Cluster & E/S0          &  $<0.071$  &  $<0.017$ \\
Cluster & S0a/b        &   0.17$^{+0.19}_{-0.10}$  &  $0.061^{+0.068}_{-0.036}$   \\
Cluster & Sbc/d        &  0.87$^{+0.41}_{-0.29}$   &  $0.610^{+0.290}_{-0.206}$   \\
Field & Total            &   0.44$^{+0.08}_{-0.07}$   &  $0.174^{+0.032}_{-0.027}$  \\
Field & E/S0             &  $<0.08$  &  $<0.020$ \\
Field & S0a/b           &  0.36$^{+0.14}_{-0.10}$   &  $0.130^{+0.049}_{-0.037}$  \\
Field & Sbc/d           &  0.83$^{+0.21}_{-0.17}$   &  $0.652^{+0.164}_{-0.134}$  \\
\hline
\multicolumn{4}{l}{Li et al. (2011b)} \\
\hline
Field & E        &   $<0.014$   & $<0.003$ \\
Field & S0      &   0.200$^{+0.015}_{-0.09}$ (0.006)   & $0.005^{+0.004}_{-0.002}$ (0.001) \\
Field & Sab    &   0.266$^{+0.047}_{-0.041}$ (0.098)   & $0.098^{+0.018}_{-0.015}$ (0.035) \\
Field & Sb      &   0.282$^{+0.043}_{-0.037}$ (0.106)   & $0.144^{+0.023}_{-0.020}$ (0.055) \\
Field & Sbc    &   0.466$^{+0.058}_{-0.052}$ (0.134)   & $0.355^{+0.042}_{-0.038}$ (0.098) \\
Field & Sc      &   0.649$^{+0.088}_{-0.078}$ ($^{+0.364}_{-0.137}$)   & $0.547^{+0.075}_{-0.066}$ ($^{+0.245}_{-0.112}$) \\
Field & Scd    &   0.795$^{+0.097}_{-0.086}$ ($^{+0.386}_{-0.135}$)   & $0.767^{+0.116}_{-0.102}$ ($^{+0.342}_{-0.154}$)
\enddata
\tablenotetext{a}{SNuB $\equiv$ SNe(100 yr $10^{10}$ $L_{B,\odot}$)$^{-1}$}
\tablenotetext{b}{SNuM $\equiv$ SNe(100 yr $10^{10}$ $M_{\odot}$)$^{-1}$}
\end{deluxetable*}

\begin{deluxetable*}{lcccccccccc}[h]
\tablecolumns{9}
\tablecaption{MENeaCS Type II Cluster Supernovae Spectroscopy Summary \label{table:CC}}
\tablehead{
\colhead{MENEACS ID} &\colhead{UT Date} & \colhead{Telescope/} & \colhead{Cluster} & \colhead{Galaxy} & \colhead{SNID} & \colhead{SNID Template} & \colhead{SNID} &\colhead{Exposure} \\
\colhead{} & \colhead{}&\colhead{Instrument}&\colhead{$z$}&\colhead{$z$}&\colhead{$z$}&\colhead{Type, Name}&\colhead{Phase ($\sigma$)}&\colhead{Time}\\
\colhead{} & \colhead{}&\colhead{}&\colhead{}&\colhead{}&\colhead{}&\colhead{}&\colhead{days}&\colhead{seconds}}\\
\startdata
Abell119\_5\_24\_0    &  2009-09-19.29  &  MMT/BCS    & 0.044 & 0.0480  & 0.050 (0.005)      &  IIP, SN92H    & 39.5 (88.0)   & 1200.0 \\ 
Abell1795\_8\_08\_1  &  2009-06-15.27  &  MMT/BCS    & 0.063 & 0.0626  & 0.0636 (0.0054)  &  IIP, SN04et   & 21.4 (55.8)   & 900.0  \\
Abell399\_11\_19\_0  &  2009-12-19.27  &  MMT/Hecto  & 0.072 & 0.072    & 0.0676 (0.0025)  &  IIP, SN04et    & 60.1 (19.2)  & 3600.0 \\
Abell1651\_7\_05\_3  &  2009-06-15.20  &  MMT/BCS    & 0.085 & ...          & 0.0739 (0.0050)  &  IIP, SN04et   & 26.9 (20.9)  & 900.0  \\
ZwCl0628\_7\_08\_0  &  2009-09-19.49  &  MMT/BCS    & 0.081 & ...          & 0.0759 (0.0050)  &  IIP, SN04et   & 48.2 (37.4)  & 900.0 \\
Abell2443\_5\_19\_0  &  2009-06-15.42  &  MMT/BCS    & 0.108 & 0.1106  & 0.1053 (0.0056)  &  IIP, SN99em & 24.0 (15.5)   & 2400.0 \\
Abell990\_6\_13\_0    &  2009-03-16.11  &  MMT/Hecto  & 0.144 & 0.1425  & 0.1422 (0.0039)  &  IIP, SN04et   & 14.3 (10.8)   & 3600.0 
\enddata
\end{deluxetable*}

\begin{deluxetable*}{lccccclc}[h]
\tablecolumns{8}
\tablecaption{MENeaCS Type II Cluster Supernovae Host Properties \label{table:hosts}}
\tablehead{
\colhead{MENEACS ID} &\colhead{$\alpha$} & \colhead{$\delta$} & \colhead{$r$} & \colhead{$g-r$} & \colhead{$\Delta (g-r)_{RS}$} & \colhead{$\rm R_{clus}$} & \colhead{$\rm R_{clus}$} \\
\colhead{} &\colhead{J2000.0}&\colhead{J2000.0}&\colhead{mag}&\colhead{mag} & \colhead{mag}&\colhead{kpc}&\colhead{$\rm R_{200}$}}\\
\startdata
Abell119\_5\_24\_0 & 00:55:39.69 & -00:52:35.9 & $16.53\pm0.02$ & $0.40\pm0.04$ & $-0.41$ & 1280 & 0.78 \\
Abell1795\_8\_08\_1 & 13:48:38.56 & +26:22:18.5 & $20.23\pm0.02$ & $0.31\pm0.04$ & $-0.37$ & 980 & 0.46 \\
Abell399\_11\_19\_0 & 02:57:16.58 & +13:08:34.3 & $16.28\pm0.02$ & $0.85\pm0.04$ & $-0.02$ & 920 & 0.49 \\
Abell1651\_7\_05\_3 &12:57:23.39 & -04:33:52.7 & $22.26\pm0.06$ & $0.22\pm0.10$ & $-0.45$ & 3530 & 1.73 \\
ZwCl0628\_7\_08\_0 & 06:31:03.31 & +24:49:17.8 & $16.11\pm0.02$ & $0.79\pm0.05$ & $-0.15$ & 1170 & 0.73 \\
Abell2443\_5\_19\_0 & 22:25:07.59 & +17:35:00.9 & $17.35\pm0.02$ & $0.56\pm0.04$ & $-0.43$ & 2350 & 1.47 \\
Abell990\_6\_13\_0 & 10:23:18.58 & +49:05:17.8 & $20.97\pm0.03$ & $0.32\pm0.05$ & $-0.63$ & 730& 0.39 
\enddata
\end{deluxetable*}

\begin{deluxetable*}{llccc}
\tablecolumns{5}
\tablecaption{MENeaCS Cluster SN\,II Rates \label{table:SNrates}}
\tablehead{ 
\colhead{Galaxy} & \colhead{Radius} & \colhead{$\rm N_{SN}$} & \multicolumn{2}{c}{SN\,II Rates}  \\
\colhead{Types}   & \colhead{}            & \colhead{}            & \colhead{SNuB\tablenotemark{a}}     &  \colhead{SNuM\tablenotemark{b}}
}
\startdata
\multicolumn{5}{l}{With the SN\,II LF of Li et al. (2010).}\\
\hline
All & 1 Mpc & 3 & $\rm 0.137_{-0.092}^{+0.318}(stat)_{-0.018}^{+0.024}(sys)$ & $\rm 0.032_{-0.020}^{+0.074}(stat)_{-0.004}^{+0.006}(sys)$ \\
All & $\rm R_{200}$ & 4 & $\rm 0.093_{-0.067}^{+0.235}(stat)_{-0.012}^{+0.063}(sys)$ & $\rm 0.026_{-0.018}^{+0.075}(stat)_{-0.003}^{+0.010}(sys)$ \\
RS & 1 Mpc & 1 & $\rm 0.044_{-0.044}^{+0.132}(stat)_{-0.004}^{+0.005}(sys)$ & $\rm 0.011_{-0.011}^{+0.035}(stat)_{-0.001}^{+0.001}(sys)$ \\
RS & $\rm R_{200}$ & 1 & $\rm 0.028_{-0.028}^{+0.061}(stat)_{-0.003}^{+0.061}(sys)$ & $\rm 0.007_{-0.007}^{+0.014}(stat)_{-0.001}^{+0.009}(sys)$ \\
Off RS & 1 Mpc & 2 & $\rm 0.346_{-0.346}^{+0.968}(stat)_{-0.045}^{+0.061}(sys)$ & $\rm 0.129_{-0.129}^{+0.443}(stat)_{-0.017}^{+0.023}(sys)$ \\
Off RS & $\rm R_{200}$ & 3 & $\rm 0.220_{-0.220}^{+0.499}(stat)_{-0.029}^{+0.362}(sys)$ & $\rm 0.080_{-0.064}^{+0.179}(stat)_{-0.013}^{+0.079}(sys)$ \\
\hline
\multicolumn{5}{l}{With the SN\,II LF of Cappellaro et al. (1993a).}\\
\hline
All & 1 Mpc & 3 & $\rm 0.112_{-0.073}^{+0.931}(stat)_{-0.015}^{+0.020}(sys)$ & $\rm 0.035_{-0.025}^{+0.231}(stat)_{-0.005}^{+0.006}(sys)$ \\
All & $\rm R_{200}$ & 4 & $\rm 0.088_{-0.063}^{+0.665}(stat)_{-0.011}^{+0.055}(sys)$ & $\rm 0.026_{-0.019}^{+0.168}(stat)_{-0.004}^{+0.010}(sys)$ \\
RS & 1 Mpc & 1 & $\rm 0.035_{-0.035}^{+0.165}(stat)_{-0.003}^{+0.004}(sys)$ & $\rm 0.008_{-0.008}^{+0.041}(stat)_{-0.001}^{+0.001}(sys)$ \\
RS & $\rm R_{200}$ & 1 & $\rm 0.023_{-0.023}^{+0.154}(stat)_{-0.002}^{+0.049}(sys)$ & $\rm 0.005_{-0.005}^{+0.047}(stat)_{-0.000}^{+0.012}(sys)$ \\
Off RS & 1 Mpc & 2 & $\rm 0.305_{-0.305}^{+2.102}(stat)_{-0.040}^{+0.054}(sys)$ & $\rm 0.106_{-0.106}^{+0.937}(stat)_{-0.014}^{+0.019}(sys)$ \\
Off RS & $\rm R_{200}$ & 3 & $\rm 0.194_{-0.194}^{+0.994}(stat)_{-0.028}^{+0.274}(sys)$ & $\rm 0.067_{-0.067}^{+0.518}(stat)_{-0.010}^{+0.071}(sys)$
\enddata
\tablenotetext{a}{SNuB $\equiv$ SNe(100 yr $10^{10}$ $L_{B,\odot}$)$^{-1}$}
\tablenotetext{b}{SNuM $\equiv$ SNe(100 yr $10^{10}$ $M_{\odot}$)$^{-1}$}
\end{deluxetable*}

\begin{deluxetable*}{llccc}
\tablecolumns{5}
\tablecaption{MENeaCS Cluster Star Formation Rates \label{table:SFR}}
\tablehead{ 
\colhead{Galaxy} & \colhead{$\rm N_{SN}$} & \colhead{$\rm R_{II}$} & \colhead{SFR} & \colhead{sSFR} \\
\colhead{Types}   & \colhead{}   & \colhead{$\rm SN \ yr^{-1}$}   & \colhead{$\rm M_{\odot} \ yr^{-1}$}   & \colhead{$\rm M_{\odot} \ yr^{-1} \ (10^{12}M_{\odot})^{-1}$} 
}
\startdata
All & 4 & $\rm 0.308_{-0.184}^{+0.912}$  & $\rm 56.6_{-33.7}^{+167.3} (stat) \pm 10.2 (sys)$  & $\rm 5.1_{-3.1}^{+15.8} (stat) \pm 0.9 (sys)$ \\
RS & 1 & $\rm 0.072_{-0.032}^{+0.155}$  & $\rm 13.3_{-5.8}^{+28.4} (stat) \pm 2.4 (sys)$  & $\rm 2.0_{-0.9}^{+4.2} (stat) \pm 0.4 (sys)$ \\
Off RS & 3 & $\rm 0.225_{-0.131}^{+0.466}$  & $\rm 41.3_{-24.1}^{+85.6} (stat) \pm 7.4 (sys)$  & $\rm 16.6_{-9.6}^{+39.5} (stat) \pm 3.0 (sys)$ 
\enddata
\end{deluxetable*}

\clearpage

\clearpage 

\begin{appendices}

\section{Derivation of The Inclination Correction}

A lower $SNR_{II}$ has been observed in high inclination spiral galaxies relative to low inclination spirals (e.g. Cappellaro et al. 1993b; Cappellaro et al. 1999). This difference is attributed to dust obscuring SNe when our line of site goes through the disk. It was suggested by Cappellaro et al. (1993b) that this only affects photographic surveys. However, Li et al. (2011b) is a CCD survey and they report a strong trend between $SNR_{II}$ and inclination for late-type spirals, but a very mild trend for early-type spirals.

Although the SN\,II LF of Li11a was not corrected for host extinction, using it does not mean we have included an inclination correction factor. For galaxy targeted SN surveys the inclination of each spiral galaxy is inferred from its shape, and the inclination correction derived from the observed rates. However, without redshifts for all galaxies in the CFHT MegaCam field of view, we cannot identify which spirals belong to the cluster and our inclination correction must be statistically estimated. 

To derive our inclination correction factor, we first describe the simple case of having two samples of galaxies, ``regular" and ``inclined". The fraction of SN\,II host galaxies which are ``regular" is $f_{H(r)}$, the fraction which are ``inclined" is $f_{H(i)}$, and the fraction of all occurring SNe\,II that {\it are} detected is $f_{SN_d(r)}$ and $f_{SN_d(i)}$. The inclination-corrected number of SNe\,II, $N_{II,c}$, given the total number observed, $N_{II}$,  is $N_{II,c} = C_{inc} N_{II}$ where the inclination correction factor $C_{inc}$ is:

\begin{equation}
C_{inc} = f_{H(r)} \frac{1}{f_{SN_d(r)}} + f_{H(i)} \frac{1}{f_{SN_d(i)}} .
\end{equation}

Instead of simply ``regular" and ``inclined", we consider 7 different galaxy types: ellipticals, and spirals of early (Sab) or late (Scd) types which are face-on (f, $0<i<40$), inclined (i, $40<i<75$), or edge-on (e, $75<i<90$). From our observations we know the approximate fractions of SNe\,II hosted by elliptical and spiral galaxies are $f_{HE}\sim1/7$ and $f_{HS}\sim6/7$. Based on the ratio between LOSS rates in field Sab and Scd spiral galaxies listed in Table \ref{table:IIrates}, SNe\,II occur $\sim10$ times more often in Scd than Sab spirals. We break $f_{HS}$ down into components $f_{HSab}=0.086$ and $f_{HScd}=0.771$. Maltby et al. (2011) find he number fraction of cluster spirals which are face-on, inclined, and edge-on is 0.16, 0.66, and 0.18 respectively. The fractions of SN\,II hosts in each of our 7 considered galaxy types are:
$f_{HE}=0.143$, 
$f_{HSab(f)}=0.014$,
$f_{HSab(i)}=0.057$,
$f_{HSab(e)}=0.015$,
$f_{HScd(f)}=0.0123$,
$f_{HScd(i)}=0.509$, and
$f_{HScd(r)}=0.139$.

Assuming all SNe\,II in face-on spirals are detected, we derive the fraction of SNe\,II detected in inclined and edge-on spirals from the rates presented in Table 3 of Li et al. (2011b). MENeaCS and LOSS are both multi-epoch CCD SN surveys at low redshift, and it is reasonable to expect a similar inclination effect. LOSS did not target galaxies expected to produce extremely extincted transients such as very dusty starburst galaxies, but neither does MENeaCS since starbursts are not generally found in rich galaxy clusters. They consider the same 7 galaxy types as described above, and from their work we find $f_{SN_d(E)}$=1.0, $f_{SN_d(Sab(f))}$=1.0, $f_{SN_d(Sab(i))}$=0.78, $f_{SN_d(Sab(e))}$=0.6, $f_{SN_d(Sab(f))}$=1.0, $f_{SN_d(Sab(i))}$=0.63, $f_{SN_d(Sab(e))}$=0.32. This results in a MENeaCS inclination correction factor of $C_{incl}$=1.62. By varying the component terms between their minimum and maximum values (e.g. the LOSS rates uncertainties), we estimate a $\sim5$\% uncertainty on $C_{inc}$ and add this to our systematic uncertainties. Note that $C_{inc}$ does not apply to our rates in red sequence galaxies.

\section{Approximation to a Rate Incorporating an ``Effective" Control Time}

The occurrence of unphysical rates can be avoided if, instead of choosing randomly from the SN\,II LF for each realization of the Monte Carlo, an effective control time is calculated using a sum over the components of the LF. In this method, Equation \ref{e:rate} becomes:

\begin{equation} \label{e:rate_v2}
SNR_{II} = \frac{C_{inc}}{C_{spec}} \ \sum_i f_i \cdot \frac{N_{II,i}}{\sum_j \Delta t_{ij} M_j} ,
\end{equation}

\noindent
where $f_i$ is the weight of the $i^{th}$ bin of the SN\,II LF, $N_{II,i}$ is the number of observed SNe\,II in the $i^{th}$ bin, and $\Delta t_{ij}$ is the control time for SN\,II from the $i^{th}$ bin. This method is derived in the Appendix of Leaman et al. (2011), and is used by LOSS. Since MENeaCS light curves are sparsely sampled, we do not have $N_{II,i}$, and this method is not appropriate for our dataset. We implement an approximation to this method by taking $N_{II}$ outside the sum over $i$, and bringing $\sum_i f_i$ inside $\sum_j$ to calculate an ``effective" control time, integrated over the SN\,II LF, for every epoch. This style of rates calculation is more processor intensive, and for this trial we only run 100 realizations. The resulting rate for ``All" galaxy types within $R_{200}$ is $SNR_{II} = 0.023_{-0.015}^{+0.030}$ SNuM, and no unphysical rates are encountered during the Monte Carlo. The upper uncertainties are smaller than those for our main method, $SNR_{II}=0.026_{-0.018}^{+0.075}$, because the SN\,II LF does not contribute to the width of the rates distribution from the Monte Carlo realizations. We reiterate the subtle point that this is not appropriate for a survey such as MENeaCS in which the observed LF, $N_{II,i}$, is not well characterized.

\end{appendices}

\end{document}